\documentclass[%
 superscriptaddress,
 preprint,
 amsmath,amssymb,
 aps,
 prb,
]{revtex4-2}

\usepackage{graphicx}
\usepackage{dcolumn}
\usepackage{bm}
\usepackage{xcolor}

\begin{document}


\title{Scanning through space and time: past, present, and future of time-resolved scanning transmission soft X-ray microscopy}

\author{Simone Finizio}
\email{simone.finizio@psi.ch} 
\affiliation{Swiss Light Source, Paul Scherrer Institut, 5232 Villigen PSI, Switzerland}

\author{Tim A. Butcher}%
\affiliation{Max Born Institute for Nonlinear Optics and Short Pulse Spectroscopy, 12489 Berlin, Germany}

\author{Sebastian Wintz}%
\affiliation{Helmholtz Zentrum Berlin f\"ur Materialien und Energie, 12489 Berlin, Germany}

\author{Markus Weigand}%
\affiliation{Helmholtz Zentrum Berlin f\"ur Materialien und Energie, 12489 Berlin, Germany}

\author{J\"org Raabe}
\affiliation{Swiss Light Source, Paul Scherrer Institut, 5232 Villigen PSI, Switzerland}

\date{\today}

\begin{abstract}

Time-resolved microscopy with the pump-probe protocol is one of the most important techniques for the investigation of dynamical processes at the nanoscale, thanks to the possibility of combining nanometric resolution imaging with sub-nanosecond temporal resolutions. Amongst the ensemble of time-resolved microscopy techniques, time-resolved scanning transmission X-ray microscopy has been, since its inception in 2006, extensively utilized for the study of magneto-dynamical processes. In this review, an overview of the concept and experimental implementations of the pump-probe protocol in time-resolved scanning transmission X-ray microscopy imaging will be presented together with some examples of recent applications of the technique. Possible future developments aimed at meeting the new opportunities and challenges offered by the upgrade of synchrotrons to diffraction limited lightsources will also be discussed.

\end{abstract}

\maketitle


\section{Introduction - Time-resolved X-ray microscopy}

In this review paper, we will concentrate on dynamical processes occurring at the microscopic scale (10$^{-6}$--10$^{-9}$ m) and at the MHz to several GHz frequency range (10$^{-6}$--10$^{-10}$ s). These length- and timescales encompass a rich ensemble of dynamical processes in condensed matter physics, including e.g., spinwave or magnon dynamics \cite{Wintz2016, art:albisetti_spinWaveOptics, Girardi2024, Traeger2021}, the dynamics of magnetic domains and domain walls \cite{art:baumgartner_switching, art:finizio_CIDWM, Bisig2013, Rhensius2010, Raabe2005}, topological solitons \cite{Litzius2017, Buettner2015, art:finizio_SkyrmionNucleation, Nagaosa2013, Woo2016, Woo2018}, and stochastic processes \cite{Zazvorka2019, Klose2023}. The experimental investigation of these processes is strongly intertwined with their dynamics, i.e., a non-invasive experimental technique able to resolve changes in the material of interest at the nanometric scale combined with the possibility to visualize the changes in the state of the sample at short timesteps is necessary. The only set of techniques meeting these requirements is time-resolved (magnetic) microscopy.

In particular, there are three main requirements for the microscopy technique. The first is that the process one wishes to image should be able to be visualized, i.e., a \textit{contrast mechanism} is needed. For many of the processes introduced above, monochromatic X-rays with different polarizations (linear and circular) provide several useful contrast mechanisms. The first is given by the element-dependent absorption of monochromatic X-rays: an element exhibits a higher absorption cross section when the X-rays are tuned to one of the elemental absorption edges \cite{Henke1993}. The second useful contrast mechanism is given when the interaction of a polarized X-ray beam with the sample is investigated. If tuned to the appropriate elemental edge of a magnetic element (e.g., the L$_3$ edge of Fe), the absorption cross section of circularly polarized X-rays is  proportional to the dot product between the wave vector of the incoming X-ray beam and the local orientation of the magnetization vector of the magnetic element. This effect is called X-ray magnetic circular dichroism (XMCD) \cite{Schuetz1987}. A similar class of dichroism, collectively falling under the name of X-ray linear dichroism (XLD) \cite{Ade1993, Stoehr1999}, occurs if the X-ray beam is linearly polarized. The changes in the absorption cross sections depend e.g., on the ferroelectric orientation of a domain with respect to the electric field axis of the X-ray beam and this can be used to probe the orientation of these ferroic domains.

The second requirement is that the microscopy technique should be able to spatially resolve the features to be investigated. Due to their nanometric wavelengths, X-rays are a suitable probe for processes occurring at the nanoscale, allowing one to transcend the limits imposed by the longer wavelengths of techniques based on visible-light. Various direct imaging X-ray microscopy techniques, both full-field (e.g., photoemission electron microscopy - PEEM \cite{Schoenhense1999}, full-field transmission X-ray microscopy - TXM \cite{Fischer1996}) and scanning (e.g., scanning transmission X-ray microscopy - STXM \cite{Kortright2000}), exist. Depending on the specific experimental requirements, one or more of these techniques can be selected.

The third requirement is the focus of this review: a mechanism for the acquisition of a \textit{timetrace} of the dynamical process of interest. At the time of writing, no \textit{real-time} imaging technique able to capture a full dynamical movie of a process at the ns to \textmu s scale with sub-ns temporal resolution exists, due to the extreme demands this would put both on the detector and on the sample itself. However, if the dynamical process of interest is \textit{fully reproducible} and \textit{periodic}, this can be used to probe the dynamical process over several cycles of the excitation, reducing the demands on both the detection and the sample to feasible levels. This method, described in detail below, is called \textit{pump-probe} imaging. 

This review will describe the pump-probe protocol and its specific implementation for time-resolved STXM imaging. While this technique has been extensively utilized for the investigation of (magneto-)dynamical processes in condensed matter physics, it is by far not the only available (time-resolved) magnetic microscopy technique. For sake of completeness, we mention other techniques that can be used for the investigation of magnetic systems at the nanoscale \cite{Christensen2024} and provide literature references for additional details thereof. The non-comprehensive list includes magnetic force microscopy (MFM) \cite{Martin1987, Koblischka2003}, scanning tunneling microscopy (STM) \cite{Wiesendanger2001}, transmission electron microscopy (TEM) \cite{Hale1959, Chapman1999}, also with the possibility to perform time-resolved investigations \cite{Wessels2022}, Kerr microscopy \cite{McCord2015}, again with the possibility to perform time-resolved investigations \cite{Hiebert1997, Freeman1996}, nitrogen vacancy (NV) microscopy \cite{Maletinsky2012, Maertz2010} and its time-resolved imaging variant \cite{Herb2025}, PEEM \cite{Schoenhense1999, Stoehr1993}, and time-resolved variants \cite{Raabe2005, Buess2007, LeGuyader2012}, coherent diffractive imaging (CDI) \cite{Eisebitt2004, Battistelli2024}, including time-resolved holography \cite{vonKorffSchmising2014, Buettner2015}, full-field TXM \cite{Fischer1996}, with one of the pioneering implementations of time-resolved magnetic microscopy more than 20 years ago \cite{Stoll2004}, and several other techniques \cite{Christensen2024}. Although outside of the scope of this review, it should be mentioned here that there is also ongoing work using full-field imaging aimed at studying \textit{non reproducible} processes such as e.g., the thermal hopping of magnetic domain walls between pinning sites \cite{Klose2023}.

\section{Pump-probe protocol}
\label{sec:pumpProbe}

The pump-probe protocol is a well-established experimental method for the investigation of dynamical processes since half a century \cite{Shank1979}. In its classical implementation, two synchronized periodic signals are required: a pump signal, used to excite the desired dynamical process (e.g., a radio frequency - RF - electrical excitation, a laser pulse, etc.), and a probe signal (e.g., an X-ray pulse, an optical pulse, etc.), which is used to probe the status of the sample at a given time instant in the pump cycle. By changing the delay in time between the pump and the probe signal, a time series of the dynamical processes that occur within the sample can be obtained.

One important consideration, true for all pump-probe imaging protocols, is that the recorded timetrace requires \textit{multiple} iterations of the dynamical process to be recorded, both in the acquisition of the state of the sample at a given delay point (in normal cases, a few million repetitions of the dynamical process are necessary for the acquisition of the sample state - see e.g., \cite{Bisig2013}), and in the fact that several measurements need to be performed for the acquisition of the full timetrace (see e.g., \cite{Buettner2015, Donnelly2020}).

For X-ray techniques, the probe signal is an X-ray pulse, such as e.g., those generated by a synchrotron lightsource. Due to their circular geometry, synchrotrons are an intrinsically pulsed X-ray source, potentially delivering one X-ray pulse at every cycle of the frequency at which the RF accelerator cavities are operated. As synchrotron lightsources offer a set of different filling patterns, i.e. how much charge the single electron bunches are filled with, there are several possible implementations of the pump-probe protocol, also dependent on the available detectors. In the following, a set of pump-probe concepts will be described. The description will focus on the methodological concept independently of its practical implementations, which will instead be described in more detail in sections \ref{sec:past} and \ref{sec:present}.

\subsection{Single pump-single probe}
\label{sec:classicPP}

Let us start with the simplest, classical, implementation of the pump-probe protocol, and consider a filling pattern containing a single electron bunch (usually called single bunch - SB - filling pattern). With the SB filling pattern, an X-ray pulse is emitted at every revolution of the SB inside the storage ring. The revolution frequency is known, as this is determined by $f_\mathrm{SB} = f_\mathrm{master}/N$, where $f_\mathrm{master}$ is the frequency of the RF cavities of the storage ring (usually referred to as \textit{master clock}) and $N$ is the maximum number of available bunches inside the ring (determined by the circumference of the ring and the master clock frequency - e.g., for the Swiss Light Source - SLS - with 288 m circumference and 500 MHz master clock, 480 bunches are available, yielding $f_\mathrm{SB} = 1.0416$ MHz).

\begin{figure}[h]
    \centering
    \includegraphics[width=0.4\textwidth]{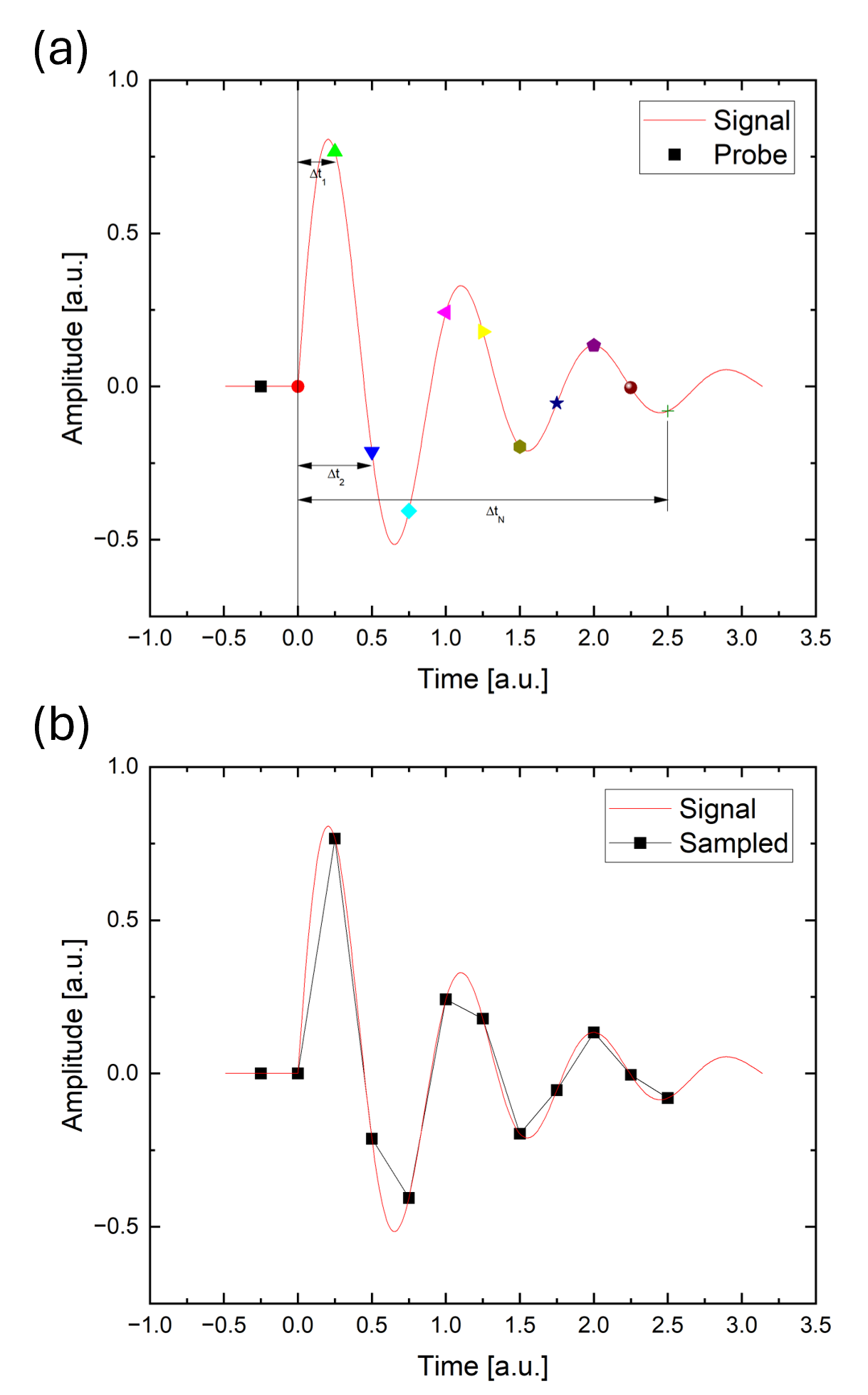}
    \caption{Sketch of the principle of pump-probe imaging. (a) A periodic signal is probed at different time delays with respect to a synchronization marker. (b) A time-trace is then reconstructed from the recorded state of the sample at the single time delays $\Delta$t$_x$. Note here that the sampled time delays need to be sufficiently dense to resolve the oscillations of the dynamical process under investigation.}
    \label{fig:pumpProbe}
\end{figure}

Let us now have a signal used to excite a dynamical process on our sample (which is the \textit{pump} signal) with a period equal to $1/(n f_\mathrm{SB})$, $n$ being an integer number, which can be either an RF signal oscillating at an integer multiple of $f_\mathrm{SB}$ or a pulsed excitation such as e.g., a laser pulse. As the two signals are frequency locked, the X-ray \textit{probe} pulse always illuminates the sample at the same instant (phase) of the pump signal. By introducing a delay between the two signals (e.g., through an optical or electrical delay line), a different phase of the excitation can be probed. By probing a set of different delays, a time series of the dynamical process can then be reconstructed, as schematically depicted in Fig. \ref{fig:pumpProbe}.

For synchrotron-based investigations, the classical pump-probe method has one major limitation, given by the requirement of a single probe pulse. SB filling patterns are associated with a delivered photon flux about 99\% lower than a regular multibunch filling pattern, which not only brings with it longer acquisition times, but is also detrimental for the other beamlines of the lightsource, where mostly quasi-static investigations are performed (i.e., SB beamtimes are rarely offered). A good compromise is to use the so-called \textit{hybrid} mode filling pattern. In this filling pattern, the majority of the electron bunches are filled. Between the filled bunches, a region with several unfilled bunches, usually called \textit{gap}, is present. Close to the center of the gap, a single bunch, called \textit{camshaft}, is filled with a charge about 3--4 times higher than the other filled bunches. By gating the detector to be sensitive only when the X-rays generated by the camshaft bunch are interacting with the sample, a time-resolved investigation with the classical pump-probe method can be performed without affecting the photon flux delivered to the other beamlines. However, the achievable measurement statistics are still the same as for a SB filling pattern, and the time series is still composed of several separate measurements at different delays.

\begin{figure}[h]
\centering
 \includegraphics[width=0.4\textwidth]{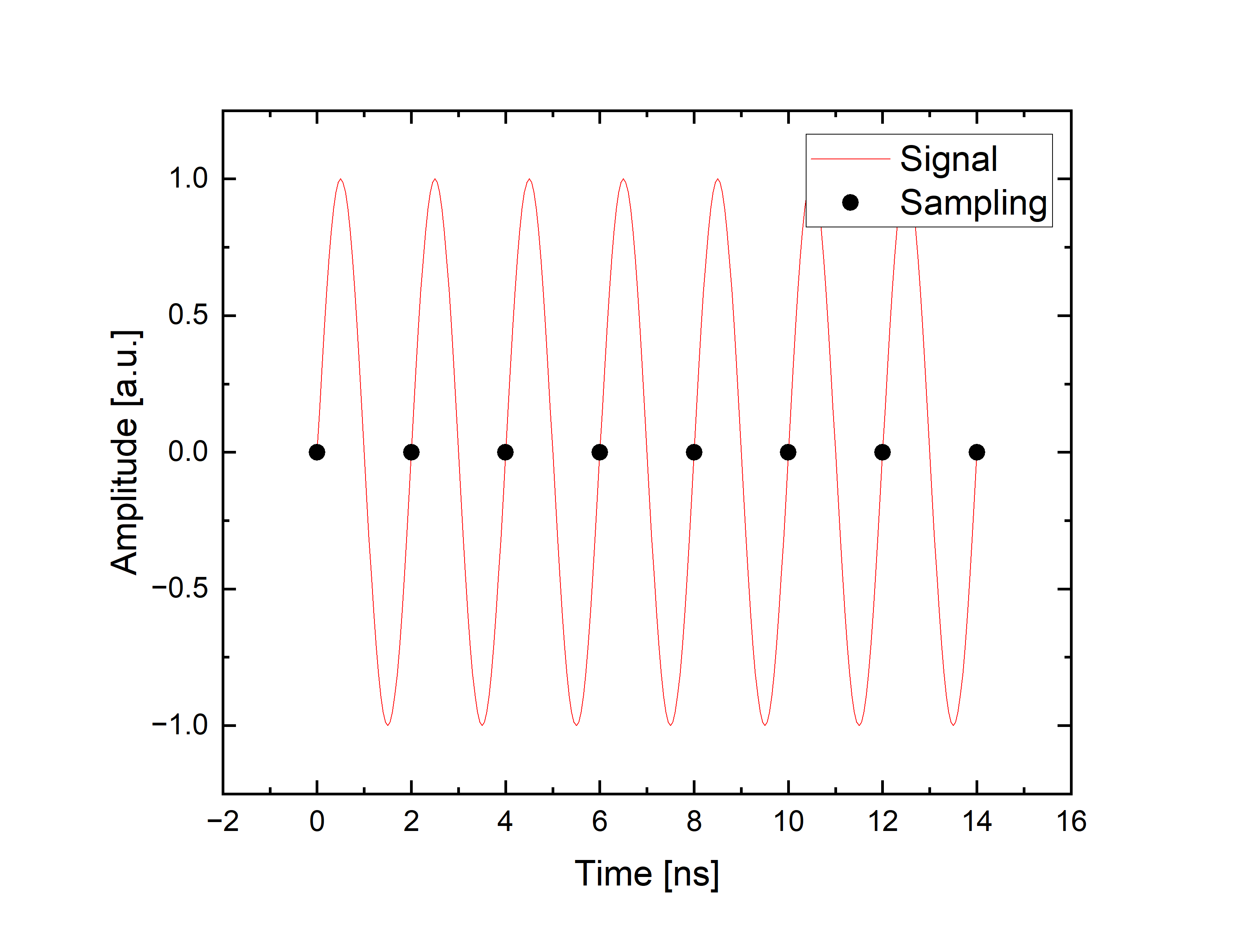}
 \caption{Special case of pump-probe imaging at synchrotron lightsources where the dynamical process under investigation is locked to the master clock frequency of the light source. In this case, all X-ray pulses probe the status of the sample at the same phase of the excitation, allowing one to significantly improve the achievable signal-to-noise ratio if compared to a SB filling pattern. As for the standard case, a time trace is then recorded by changing the delay between the excitation signal and the master clock.}
 \label{fig:500MHz}
\end{figure}

A special case occurs if the sample is excited at an integer multiple of the master clock frequency. In this case, the X-ray photons emitted from \textit{all} electron bunches illuminate the sample at the same phase. This is schematically shown in Fig. \ref{fig:500MHz}, where the 500 MHz master clock frequency of the SLS has been used as an example. This allows for the use of the entire filling pattern for probing the dynamical process excited in the sample, significantly improving the achievable signal-to-noise ratio if compared to using a SB, but severely limiting the comb of frequencies that can be investigated.

\subsection{Single pump-multiple probe - frequency locked}
\label{sec:freqLock}

The classical pump-probe protocol described in section \ref{sec:classicPP} records a time trace by performing multiple measurements in which the delay between the pump and the probe signals changes. This means that the time trace is recorded as a sequential series of different points, potentially making it vulnerable to progressive changes of the sample (even though this can be amended by acquiring a non-sequential series of delays \cite{Buettner2015, Donnelly2020}) and providing certain important experimental constraints for synchrotron-based measurements, such as either the requirement to use a SB filling pattern, or to lock the excitation to integer multiples of the master clock.

\begin{figure}[h]
\centering
 \includegraphics{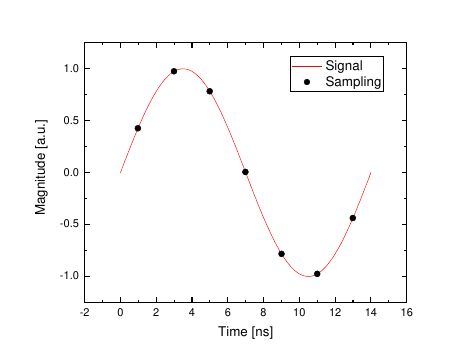}
 \caption{Example of the time-resolved sampling of a sine wave with $N=7$ and $M=1$. Seven sampling points every 2 ns are marked with black dots.}
 \label{fig:TR_STXM:RF_N7_M1}
\end{figure}

Let us now assume that we have a detection setup able to resolve X-ray photons emitted from neighboring electron bunches in the filling pattern and that we are able to ``tag'' from which electron bunch the detected photon was emitted. Let us now also apply an external excitation to our sample, leading to oscillating dynamics, with a frequency $f$ given by a fractional multiple of the master clock, i.e.:
\begin{equation}
N \cdot f = M \cdot f_\mathrm{master} , \label{eq:RF_FPGA}
\end{equation}

\noindent where $M$ indicates the number of times that the excitation signal is repeated within a period of $N/f_\mathrm{master}$. For simplicity and to provide a practical example, let us consider $f_\mathrm{master} = 500$ MHz. Therefore, an X-ray pulse illuminates the sample every 2 ns, and the absorption probability depends on the dynamical state of the sample. As the excitation is frequency locked to the X-ray pulse repetition frequency, only a finite number of points in the excitation timetrace are sampled, as shown for an example sine wave with $N=7$ and $M=1$ in Figure \ref{fig:TR_STXM:RF_N7_M1}.

\begin{figure}[h]
\centering
 \includegraphics{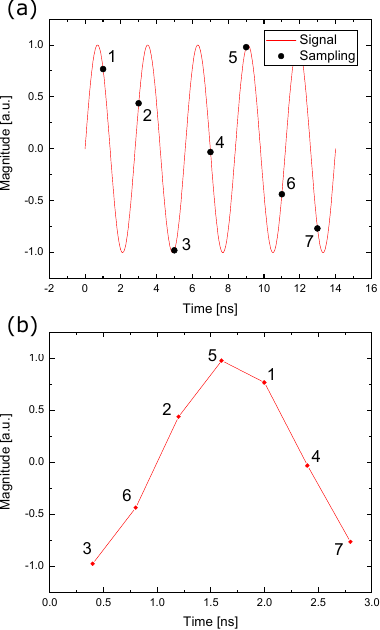}
 \caption{(a) Example of sampling a sine wave with $N=7$ and $M=5$ and the indicated sampling points. (b) Reconstructed signal, obtained from re-sorting the sampling points according to Eq. \eqref{eq:FPGA_reconstruction}.}
 \label{fig:TR_STXM:Reconstruction}
\end{figure}

Using the example shown in Fig. \ref{fig:TR_STXM:RF_N7_M1}, $N=7$ points of the excitation are sampled. As each recorded photo can be assigned to the electron bunch from which it was emitted, we can also allocate it to one of the $N$ points of the frequency-locked excitation simply by calculating $n_\mathrm{bunch} \, \mathrm{mod} \,  N$, where $n_\mathrm{bunch}$ is the bunch number from which the photon originated. However, what is shown in Fig. \ref{fig:TR_STXM:RF_N7_M1} is a special case as $M=1$, and the sampled points represent sequential phases in the RF excitation. Eq. \eqref{eq:RF_FPGA} states that there are $M$ periods of the excitation at frequency $f$ within $N$ periods of the master clock. Therefore, if $M>1$, the sampled points shown in Fig. \ref{fig:TR_STXM:RF_N7_M1} will no longer be sequential in phase, as shown exemplarily in Fig. \ref{fig:TR_STXM:Reconstruction}(a), where $N=7$ as before, but $M=5$. 

The case of $M>1$ requires the sorting of the recorded timetrace. This is done by calculating the phase of each of the sampling points with respect to the excitation signal according to the following equation:
\begin{equation}
\phi_i = 2 \pi \frac{i M (\mathrm{mod} N)}{N},
\label{eq:FPGA_reconstruction}
\end{equation}
where mod denotes the modulo function, and $i$ denotes the $i$-th sampling point. The sampling points are then sorted according to their phase. In our example of $N=7$ and $M=5$ shown in Fig. \ref{fig:TR_STXM:Reconstruction}, the reconstructed phases would be as shown in Tab. \ref{tab:TR_STXM:Reconstruction}. Reordering the sampling points according to their phase would then lead to the reconstructed time-trace shown in Fig. \ref{fig:TR_STXM:Reconstruction}(b).
\begin{table}[h]
\centering
\begin{tabular}{| r | c | c | c | c | c | c | c |}
 \hline
 \bf Sampling point & 1 & 2 & 3 & 4 & 5 & 6 & 7  \\
 \hline
 \bf Phase & 5/7 & 3/7 & 1/7 & 6/7 & 4/7 & 2/7 & 0/7 \\
 \hline
\end{tabular}
\caption{Calculated phases (normalized to $2 \pi$) for the sampling points shown in Fig. \ref{fig:TR_STXM:Reconstruction}(a).}
\label{tab:TR_STXM:Reconstruction}
\end{table}
An important consideration for this reconstruction protocol is that, by probing over different periods of the excitation signal, it is possible to reconstruct the time trace of excitations with a period below $1/f_\mathrm{master}$ whilst keeping a sampling frequency of $f_\mathrm{master}$. This is shown in the example depicted in Figure \ref{fig:TR_STXM:Reconstruction}(b), where the reconstructed signal now has a time step of 400 ps (i.e. 2/5 ns). In general, the reconstruction described above allows one to obtain a time trace of $N$ sampling points separated by a time step of $1/(f_\mathrm{master}\,M)$. It should be noted here that the time step defined above is not synonym to the achievable time resolution, which is discussed below.

\begin{figure}[h]
\centering
 \includegraphics{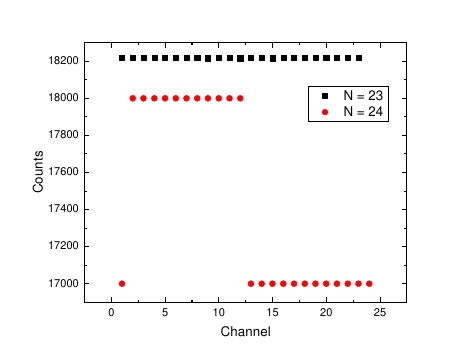}
 \caption{Calculated recorded counts for a flat signal using the SLS filling pattern (480 bunches, 420 of which filled) for a number of channels $N$ of 24 and 23. It is immediately visible that, for $N = 24$, the gap in the filling pattern gives rise to a detectable artifact, due to the fact that the X-ray bunches do not contribute equally to all of the 24 channels. This artifact disappears if $N$ is coprime with the number of electron bunches. The counts are calculated over 1000 synchrotron cycles, assuming that each filled bunch contributes 1 photon to the channel.}
 \label{fig:TR_STXM:artefacts}
\end{figure}

The single pump-multiple probe method described above is based on the assumption that the X-rays illuminate the sample at specific moments in time, separated by $1/f_\mathrm{master}$. However, in reality, there are three important considerations to take care of. The first is that the filling pattern may contain a gap. Although the machine physics reasons for this are outside the scope of this work, their consequences are not. In particular, this means that, for a certain number of bunches, there no X-rays are emitted (using the SLS as example, only 420 out of 480 bunches are typically filled \cite{Finizio2020}). However, the acquisition protocol described above assumes an equal illumination of all the $N$ sampling points of the RF signal. If this condition is not met, artifacts appear in the reconstructed time trace, as shown exemplarily in Fig. \ref{fig:TR_STXM:artefacts}. In the figure, two scenarios are shown: the first has $N$ selected to be a divisor of the total number of electron bunches, which leads to a non-uniform illumination in the $N$ sampling points. The second, instead, has $N$ selected to be coprime with the number of electron bunches. This guarantees that each of the $N$ points is sampled uniformly by all of the electron bunches, removing any artifacts arising from the filling pattern structure.

\begin{figure}[h]
\centering
 \includegraphics{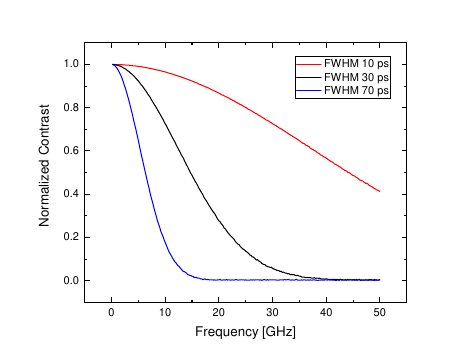}
 \caption{Calculated amplitude of recorded time traces when considering different X-ray pulse FWHMs as a function of the frequency of the time-resolved signal that is being recorded. Image from \cite{Finizio2021}.}
 \label{fig:TR_STXM:ToAErrors}
\end{figure}

The second consideration is that the X-ray arrival time is distributed over a time interval defined by the width of the electron bunches \cite{Finizio2020}. The protocol described above, however, assumes that the photons arrive at a well-defined moment in time, corresponding to the effective phases calculated in Eq. \eqref{eq:FPGA_reconstruction}. This means that the dispersion of the photon arrival time within an X-ray bunch introduces an error in the recorded signal in each of the sampling points, given by the convolution between the distribution of the photon arrival times and the signal to be recorded \cite{Finizio2021}. This limits the maximum frequency that can be detected with this method and is dependent on the X-ray pulse width, as shown in Fig. \ref{fig:TR_STXM:ToAErrors}.

 Once again using the SLS as an example, the typical width of the electron bunches is of about 70 ps full-width at half-maximum (FWHM) \cite{Finizio2020}, leading to a maximum detectable frequency of about 10 GHz (with 90\% contrast loss when compared to the quasi-static case) \cite{Finizio2021}. Note that, if the phase shift caused by the third harmonic cavities - see section \ref{sec:phaseShift} - is considered, the maximum detectable frequency will be further reduced. Some light sources such as e.g., the Bessy II synchrotron, offer a special filling pattern with low-$\alpha$ optics, with electron bunch widths in the range of 5-20 ps \cite{art:bessy_lowAlpha_VSR}. With these shorter bunches, the detectable frequency range can be extended by almost an order of magnitude compared to regular optics filling patterns \cite{Finizio2021, Weigand2022}. While low-$\alpha$ optics filling patterns enable the detection of a wider frequency range \cite{Mayr2024}, low-$\alpha$ optics beamtime is a much more restricted than regular optics beamtime (at the time of writing, only a limited set of lightsources offer it, and only for a few weeks per semester), encouraging the development of techniques that can detect excitation frequencies beyond the limit imposed by the electron bunch widths. In addition, as described in more detail in section \ref{sec:phaseShift}, there are additional factors to consider such as the shift in the phase of the electron bunches caused by the third harmonic cavities inside the synchrotron storage ring \cite{Finizio2020}, which can further reduce the detectable frequency range with the method described above.

\subsection{Single pump-multiple probe - frequency unlocked}
\label{sec:ToA}

The method described in section \ref{sec:freqLock} has two major limitations. The first, intrinsic to the approach, is a consequence of the requirement to synchronize the excitation signal to the master clock of the synchrotron light source and requires the excitation frequency to be a fractional multiple of the master clock frequency according to Eq. \eqref{eq:RF_FPGA}. The second limitation, also mentioned in section \ref{sec:freqLock}, is based on the assumption that the X-ray photons always interact with the sample at the same phase points defined according to Eq. \eqref{eq:FPGA_reconstruction}. As the electron bunches have a non-negligible width, this results in an uncertainty on the X-ray photon arrival time, i.e. on the phase of the excitation that is probed by the photon. The consequence, as detailed in section \ref{sec:freqLock} and in Fig. \ref{fig:TR_STXM:ToAErrors}, is a blurring of the recorded time trace along the time axis and a limit to the minimum detectable excitation period, which is comparable to the width of the electron bunches \cite{Finizio2021}. 

While the latter limitation can be partially amended, albeit with some compromises on the imaging statistics, by using shorter electron bunches (e.g., beamtime with low-$\alpha$ optics), the former is intrinsic to the method. However, both can be overcome if the assumption on the arrival time of the X-ray photons is lifted and the time difference between the photon arrival time and the externally-driven excitation is measured.

Let us assume that we can measure the time at which the X-ray photon interacts with the detector (see section \ref{sec:ToA_implementation} for details about the specific implementation) and that we also have a marker signal synchronized with the excitation applied to the sample available, of which the ``arrival" time is also known. 

\begin{figure}[h]
    \centering
    \includegraphics[width=0.4\textwidth]{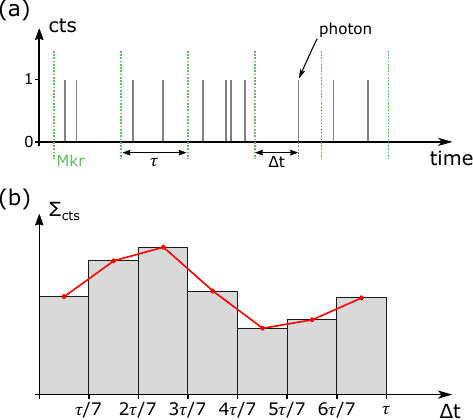}
    \caption{Sketch of the concept of time-of-arrival detection protocol. (a) In the time-of-arrival protocol, the photon arrival times are determined, yielding a set of timestamps. In parallel, timestamps of a marker signal synchronized with the excitation applied to the sample (and periodic with a period $\tau$) are also recorded. For each photon arrival time, the time-difference $\Delta$t with respect to the last recorded marker is calculated. (b) The recorded time differences are then sorted in time bins with a user-defined width (in this case, $\tau$/7), allowing one to obtain a time trace of the dynamical process. Figure from \cite{Finizio2022}.}
    \label{fig:ToA_concept_sketch}
\end{figure}

In this scenario, the recorded data would consist of a series of photon arrival timestamps with an additional set of timestamps synchronized with the applied excitation, as sketched in Fig. \ref{fig:ToA_concept_sketch}(a). The time difference between the photon timestamp ($\mathrm{\Delta}$t in Fig. \ref{fig:ToA_concept_sketch}(a)) and the last detected marker signal would then be equivalent to the time delay of a single pump-probe experiment. Therefore, a time trace can be reconstructed from the recorded time differences by sorting them as counts in user-defined time bins, as schematically shown in Fig. \ref{fig:ToA_concept_sketch}(b). 

As this method, which we dubbed \textit{time-of-arrival} detection \cite{Finizio2020}, directly measures the arrival time of the X-ray photons and compares it to a marker locked to the excitation, there are no requirements for the excitation to be locked to the master clock of the synchrotron lightsource, lifting the first limitation of the method described in section \ref{sec:freqLock}. However, this method makes the assumption that, within the acquisition time, the entire time series is sampled uniformly. In practical terms, this is achieved simply by not providing the master clock as reference clock signal to the RF setup used to generate the excitation applied to the sample. Small, unavoidable, drifts between the master clock and the internal clock of the RF setup guarantee that even selecting a fractional multiple of the master clock frequency results in a uniform sampling of the excitation.

The second limitation mentioned above is also partially lifted, as now the arrival time of the photon is measured. The determination of the temporal resolution then shifts from the uncertainty caused by the width of the electron bunch to the timing jitter of the interaction process of the X-ray photon with the diode (see section \ref{sec:LGAD} for more details) and of the electronics of the detection setup. As shown in the SLS implementation of this concept (see section \ref{sec:ToA_implementation} and \cite{Finizio2020}), this method allowed for the doubling of the detectable frequency range in normal optics operation, with still room for improvement.

\section{Intermezzo - Scanning Transmission X-ray Microscopy}
\label{sec:STXM}

The methods described in section \ref{sec:pumpProbe} above concern the detection of a time-resolved signal. However, this work reviews time-resolved \textit{scanning X-ray microscopy} and an imaging method compatible with the single pump-multiple probe protocol will also be presented. The single pump-multiple probe protocol described above sets the requirement for an X-ray detector with a sufficiently high bandwidth to resolve photons emitted from two neighboring electron bunches in the filling pattern. For the SLS, the electron bunches are separated by 2 ns (500 MHz), implying that a detector bandwidth of at least 1 GHz is desirable. At the time of writing, the only available X-ray detectors with such a high bandwidth are avalanche photodiodes (APDs). High-bandwidth APDs can be used as point detectors for X-ray photons, i.e. they have to be implemented within a \textit{scanning} microscopy technique, as they are not compatible with full-field imaging, and they need to be integrated within a photon in-photon out technique.

\begin{figure}[h]
    \centering
    \includegraphics[width=0.35\textwidth]{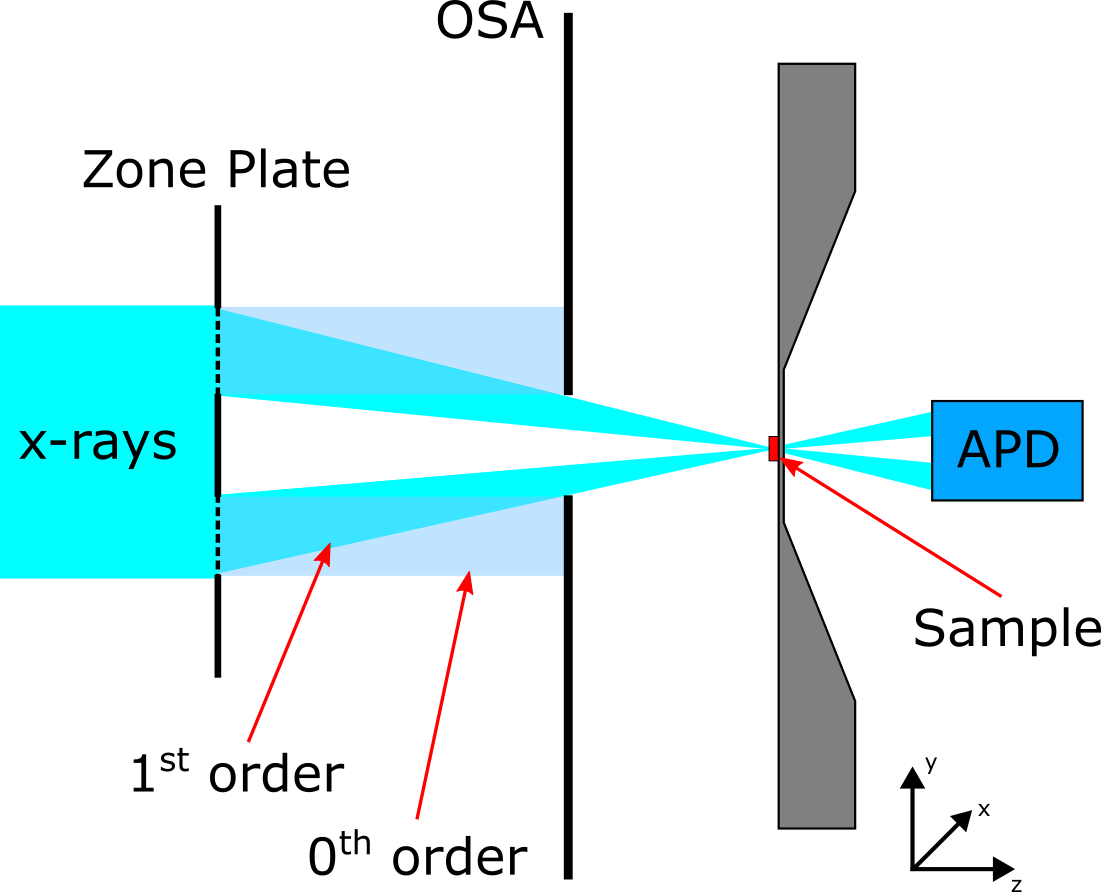}
    \caption{Sketch of the operating principle of STXM imaging. A monochromatic X-ray beam is focused by a Fresnel zoneplate onto a nanometric spot at the surface of an X-ray transparent sample. The intensity transmitted across the sample is recorded by a suitable point detector (in this case, an APD). An image is formed by scanning the sample with a piezoelectric positioner and acquiring the transmitted intensity for each point of the image. A pinhole, called order selecting aperture (OSA) in combination with a center stop mounted on the zoneplate is needed to guarantee that only the first order focus is illuminating the sample. Image from \cite{Finizio2016}.}
    \label{fig:STXM}
\end{figure}

Scanning transmission X-ray microscopy (STXM) is an implementation of a photon in-photon out X-ray microscopy technique that is compatible with the use of an APD as photon detector. As depicted in Fig. \ref{fig:STXM}, in STXM imaging a monochromatic X-ray beam delivered by the synchrotron lightsource is focused onto a nanometric spot on the surface of an X-ray transparent sample with a diffractive optical element, called Fresnel zoneplate. The intensity of the X-ray beam transmitted across the sample is then recorded with a point detector (in this case, an APD). An image is formed by scanning the sample with a piezoelectric positioner and recording the transmitted intensity for each pixel of the image. For time-resolved STXM, a time trace is also collected for each pixel of the image, allowing one to obtain a movie of the dynamical process. 

The following sections detail various practical implementations of the methods described in section \ref{sec:pumpProbe} for the acquisition of a time trace for each pixel of a time-resolved image. The approaches can be combined with all standard STXM imaging protocols, including three-dimensional imaging through STXM-laminography \cite{Witte2020}. The method for acquiring time-resolved 3D images is also described in section \ref{sec:4D}.

While the temporal resolution of a time-resolved STXM image is determined, in first instance, by the X-ray pulse width (see below for details), the achievable spatial resolution in STXM imaging is dictated, at the last instance, by the focusing optics: under a sufficiently coherent illumination of the zoneplate, the beam spot is limited by the diameter of the Airy disk generated by the zoneplate, equal to 1.22 times the outermost zone width of the zoneplate. At soft X-ray energies, the current record, limited by lithographical fabrication, is of 7 nm \cite{Roesner2020} and the usual routine value, with less stringent requirements on the experimental endstation, is of about 20 nm. While in principle out of scope for this work, a few words should be spent on the currently ongoing efforts to push the achievable spatial resolution of scanning X-ray microscopy to beyond the limits of the focusing optics. This is being carried out by developing soft X-ray ptychography imaging, which, in a simplified picture, combines scanning microscopy with coherent diffractive imaging (CDI) \cite{pfeiffer_2018}. While substantial improvements in the achievable spatial resolution in soft X-ray imaging compared to STXM have been achieved \cite{Butcher2025c, Butcher2024, Shapiro2020, Hitchcock2015}, the method requires the use of a 2D X-ray detector to record the coherent diffraction patterns generated by the interaction of the coherent X-ray beam with the sample. As described in more detail in section \ref{sec:ptychoTR}, the current technology of 2D X-ray detectors does not allow for the acquisition of time-resolved images in the single pump-multiple probe protocol. However, a reduced set of high-resolution time-resolved imaging experiments are still possible.

\section{Implementations - The past}
\label{sec:past}

The next three sections will describe the practical implementations of time-resolved STXM imaging techniques based on the single pump-multiple probe protocol. In this section, a historical overview of the first implementations, performed at the Advanced Light Source (ALS), will be presented, followed by a description of the implementations of the technique that followed at Bessy II and at the SLS.

\subsection{The first implementation}
\label{sec:als}

The first reported implementations of time-resolved STXM were performed at the ALS in the early 2000s. The experimental setups are described in detail in \cite{Acremann2007, VanWaeyenberge2006} and they are an implementation of the frequency locked method described in section \ref{sec:freqLock}.

\begin{figure*}[h]
 \includegraphics[width=0.9\textwidth]{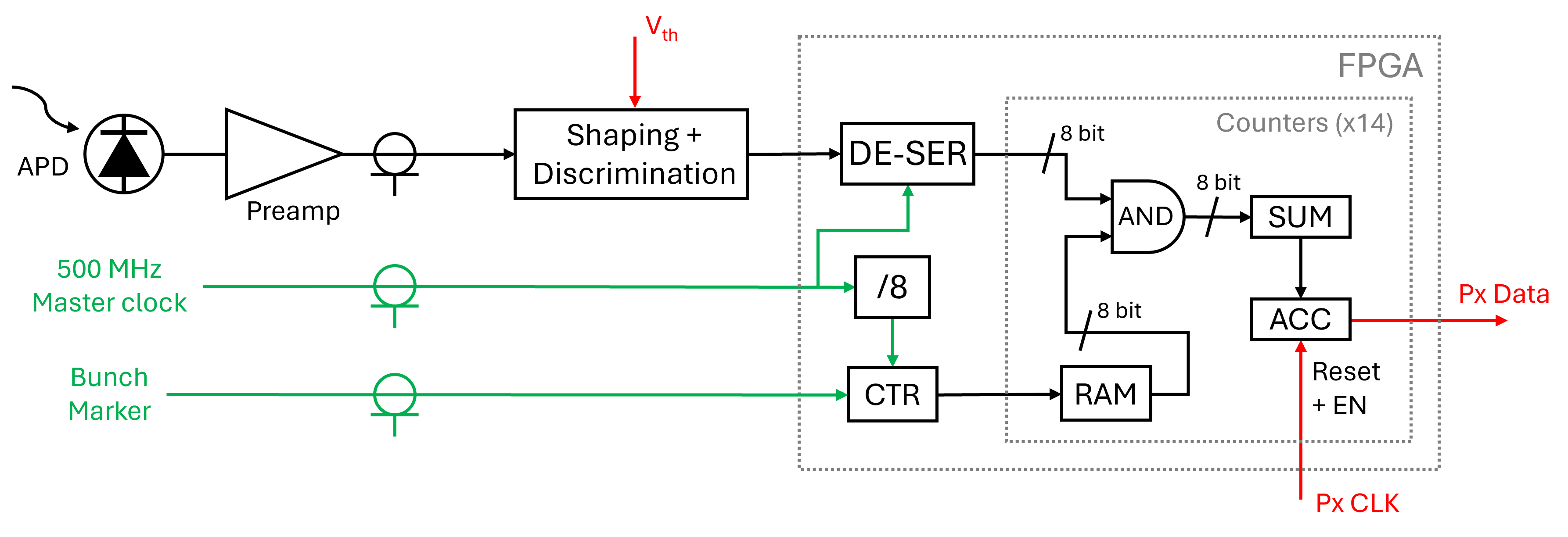}
 \caption{Simplified block diagram of the implementation of the single pump-multiple probe detection method for time-resolved STXM reported in \cite{Acremann2007}. Green lines indicate timing signals and red lines indicate external inputs/outputs. 14 counters are available in this setup.}
 \label{fig:FPGA_als}
\end{figure*}

The simplified block diagram of the setup is shown in Fig. \ref{fig:FPGA_als}. As, at the time, the hardware utilized for the setup (Altera Stratix-2 field programmable gate array - FPGA) could not directly lock to the 500 MHz master clock of the ALS and the number of memory cells available was limited and only sufficient for 14 32-bit accumulator units, a more complex logic approach had to be implemented. The FPGA was therefore driven by a 1/8 clock derived from the 500 MHz master clock (62.5 MHz), which was used to increment a memory address counter (which could be reset by a synchronization clock, e.g., synchronized to the bunch marker of the ALS \cite{Acremann2007}). The recorded data from the APD was processed by an external discriminator, generating a 500 Mbit/s signal, where a 1 identified the detection of a photon in the 2 ns window. This datastream was processed by a deserializer (DE-SER in Fig. \ref{fig:FPGA_als}), producing an 8-bit parallel signal that was updated at each cycle of the 1/8 divided clock. 

The setup was then composed of 14 independent accumulator units, each with the following components: a $8 \times 64$ bit random access memory (RAM). The RAM was written at the beginning of the measurement and recorded which electron bunches were counted in the specific accumulator unit (e.g., if the first 8 bits of the RAM were recorded with \verb|10011000|, photons originating from the 4\textsuperscript{th}, 5\textsuperscript{th}, and 8\textsuperscript{th} bunches would be counted in the accumulator unit). At each cycle of the 1/8 divided clock the de-serialized APD data (1 byte) and the selected electron bunch information (also 1 byte) contained in the RAM address selected by the counter underwent a bit-wise logical ``AND" operation. The bits in the result of the ``AND'' operation were then summed, and stored in a 32-bit accumulator. The accumulator was enabled only during the acquisition of a pixel and reset to zero at the beginning of the next pixel acquisition. The recorded data from the 14 independent counters was then transmitted to the STXM control software, where it could be combined to form the time-resolved image \cite{Acremann2007}.

Despite the limited number of accumulator units enforced by the hardware, this setup pioneered the start of time-resolved STXM imaging, allowing for the first imaging of spin-transfer torque induced switching \cite{Acremann2006} and of the magnetic field induced switching of magnetic vortex cores \cite{VanWaeyenberge2006}.

\subsection{Demultiplexer approach}
\label{sec:demux}

\begin{figure*}[h]
 \includegraphics[width=0.9\textwidth]{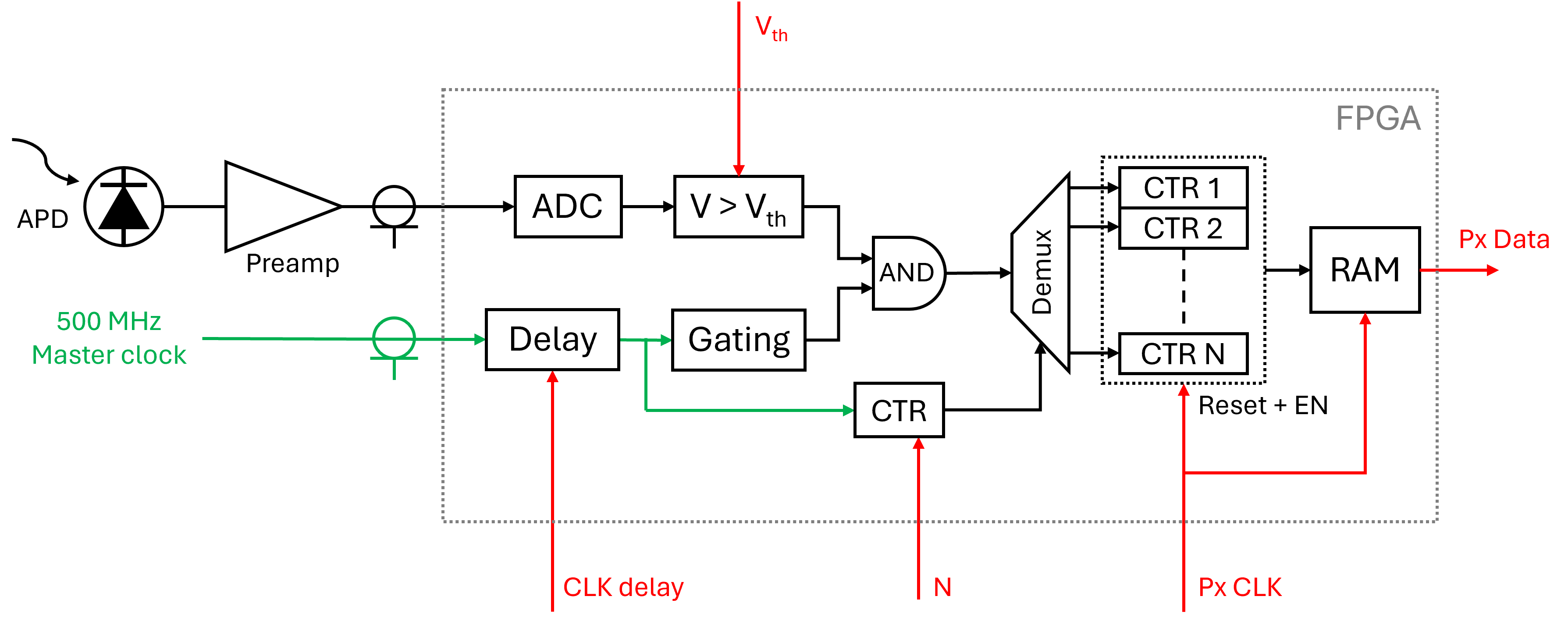}
 \caption{Simplified block diagram of the DEMUX implementation of the single pump-multiple probe detection method for time-resolved STXM. Green lines indicate timing signals and red lines external inputs/outputs. This setup requires the excitation to be locked to a fractional multiple of the master clock frequency. The maximum number of points $N$ in the recorded time series is limited and determined by the available FPGA hardware.}
 \label{fig:FPGA_demux}
\end{figure*}

This method is implemented at the Maxymus beamline of the Bessy II lightsource \cite{Weigand2022} and is also based on a FPGA setup combined with a fast analog to digital converter (ADC). This is an implementation of the frequency locked method described in section \ref{sec:freqLock}, and therefore requires an excitation with its frequency locked to a fractional multiple of the master clock.

The simplified block diagram for the setup is sketched in Fig. \ref{fig:FPGA_demux}, where the core element of the detection setup is the demultiplexer (DEMUX) used to sort the recorded photon counts. In this implementation there are several different components. The first one is the ADC, which receives the signal from the APD, with a suitable pre-amplification. Directly downstream from the APD, a voltage discriminator circuit is present. This circuit asserts 1 only if the recorded APD voltage is above a user-defined threshold (i.e., this circuit is used to discriminate whether a photon was detected by the APD). The second component handles the timing in the FPGA. Here, the master clock of the synchrotron lightsource is sampled with a user-defined delay. Downstream from this component, there are two separate components: the first, defined as ``Gating" in Fig. \ref{fig:FPGA_demux}, is used in combination with the voltage discriminator to reduce the sensitivity to external noise to the APD. The gating circuit is active only for a small period for each of the master clock cycle, and allows for the recording of a detected photon count only in the active time (achieved by a logical AND between the output of the voltage discriminator and the gating circuit). The second component is a counter that is incremented at each master clock cycle and reset once it reaches a user-defined value $N$, equal to the number of points in the time series described in section \ref{sec:freqLock}.

The value of the counter is then used to select the output of the DEMUX. A set of 16-bit counters, each connected to one of the outputs of the DEMUX, is used to record the detected photon counts in the appropriate ``bin" depending on the user-defined number of points $N$. These counters record only for the (user-defined) pixel dwell time in the time-resolved STXM image. This is carried out by enabling the counters only when the pixel clock linked to the STXM acquisition (asserted 1 when the sample is in the correct position and data should be recorded, 0 otherwise), and resetting them once the pixel has been acquired and the data stored in a random access memory (RAM), from which the recorded time trace will be transmitted to the STXM control software.

This implementation has been used for many years at the Maxymus beamline of Bessy II, and has been employed for acquiring several dynamical processes at the nanoscale (see e.g., \cite{Kammerer2011, Bisig2013, Litzius2017, Wintz2016, Woo2018, Traeger2021, Dieterle2019, Mayr2021}). Besides the limitations described in section \ref{sec:freqLock}, an additional hardware-linked limitation is present for this setup: the maximum number of points for the timetrace is restricted by the number of independent 16-bit counters that can be implemented within the FPGA used for the setup. For the Bessy II implementation, which uses an Altera Cyclone III FPGA, the maximum number for $N$ is 2048 \cite{Weigand2022}.

\subsection{RAM address approach}
\label{sec:RAM}

\begin{figure*}[h]
 \includegraphics[width=0.9\textwidth]{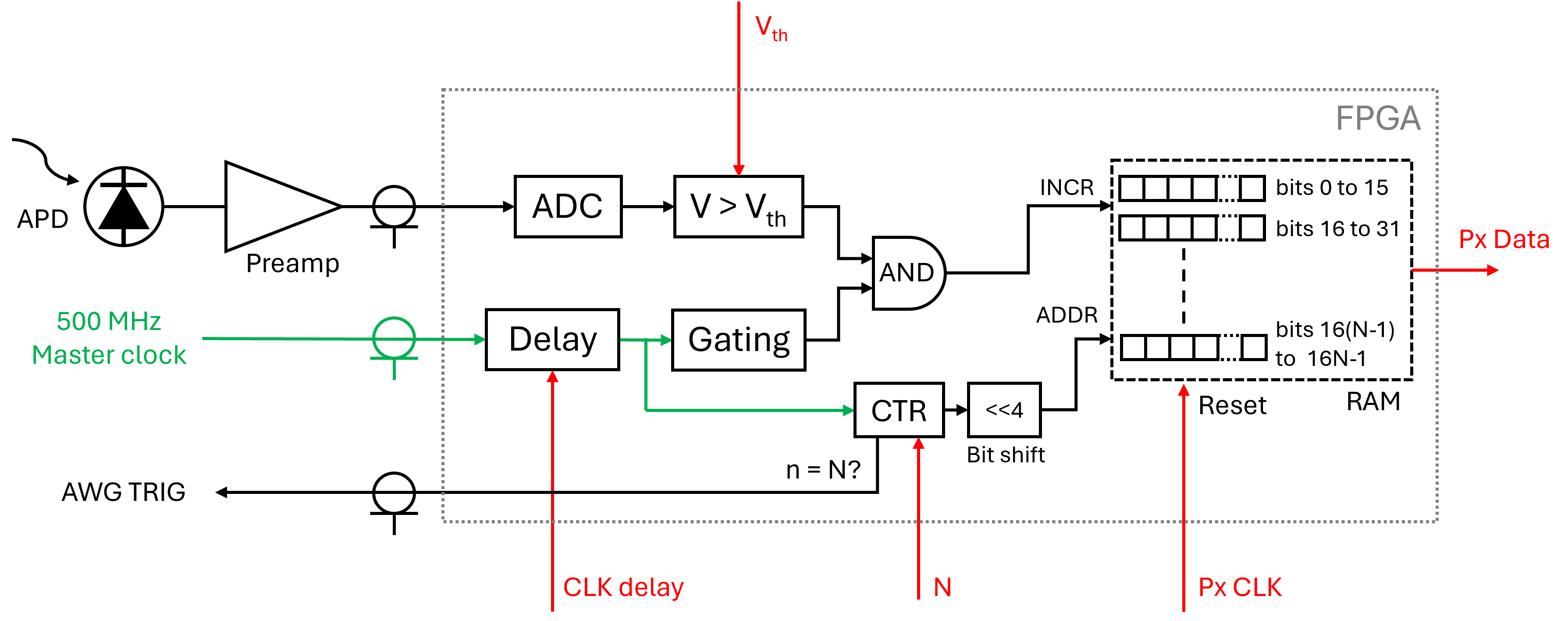}
 \caption{Simplified block diagram of the RAM address implementation of the single pump-multiple probe detection method for time-resolved STXM. Green lines signify timing signals and red lines external inputs/outputs. This setup requires the excitation to be locked to a fractional multiple of the master clock frequency. Also in this case, the maximum number of points $N$ in the recorded time series is limited and determined by the available FPGA hardware.}
 \label{fig:FPGA_RAM}
\end{figure*}

This method was implemented at the PolLux beamline of the SLS \cite{art:puzic_TR_STXM}. The operating principle (simplified block diagram shown in Fig. \ref{fig:FPGA_RAM}) is similar to the demultiplexer approach described in section \ref{sec:demux} where, instead of addressing a demultiplexer, the recorded photon counts are directly recorded in RAM cells. Upon the successful detection of a photon (i.e., APD voltage above the user-defined threshold when the gating signal is active) a 16-bit cell in the RAM is incremented by 1. The address of the cell is determined by the counter unit incremented by the master clock. As for the implementations described above, the recording is enabled only when the pixel clock is active, and the memory cells are reset once the pixel clock is inactive (of course, following the transmission of the time trace to the STXM control software).

An additional difference from the implementations described above is that the counter has a second output, asserted to 1 for the clock cycle in which the counter reaches the user-defined $N$ value. This signal is output from the FPGA setup and used as a trigger for the arbitrary waveform generator (AWG) used for the excitation of the sample. This allows one to synchronize the excitation and maintain the same time-zero value (assuming that no changes in the setup are made) even if a restart of the AWG is required.

This implementation was used for many years at the PolLux beamline of the SLS and allowed for the acquisition of several time-resolved STXM images of magnetodynamical processes, both using sinusoidal waves  (see e.g., \cite{art:albisetti_spinWaveOptics, art:finizio_MSDynamics, Girardi2024}) and arbitrary waveforms (see e.g., \cite{art:baumgartner_switching, art:finizio_CIDWM, art:finizio_SkyrmionNucleation}). The limitations of the setup are similar to the demultiplexer implementation, namely the requirement to lock the excitation to a fractional multiple of the master clock, the limitations on the achievable temporal resolution, and the hardware-linked limit on the maximum number $N$ of points, which for this setup is shifted to the available number of RAM cells. For the setup implemented at PolLux, using a Xilinx Vertex-6T FPGA, the upper limit for $N$ was of 1024, usually rounded down to the prime number 1021 \cite{art:puzic_TR_STXM}.

\section{Implementations - The present}
\label{sec:present}
The previous section reviewed the first implementations of time-resolved STXM imaging with the single pump-multiple probe protocol. This section reviews the current implementations for the beamlines that are active in time-resolved STXM imaging at the time of writing, namely the PolLux beamline of the SLS and the Maxymus beamline of the Bessy II light source, and some of the more unconventional implementations of time-resolved STXM, such as 4D imaging and the investigation of auto-oscillatory processes.

\subsection{Resampling of phase shift}
\label{sec:phaseShift}

As mentioned in sections \ref{sec:freqLock}, \ref{sec:demux}, and \ref{sec:RAM}, one assumption of the frequency locked single pump-multiple probe protocol is that the X-ray photons are emitted exactly at the center of the electron bunches. An additional implicit assumption is that the electron bunches are perfectly aligned with the master clock, i.e. that each bunch is centered at a specific phase of the master clock signal. 

\begin{figure}
    \centering
    \includegraphics[width=0.45\textwidth]{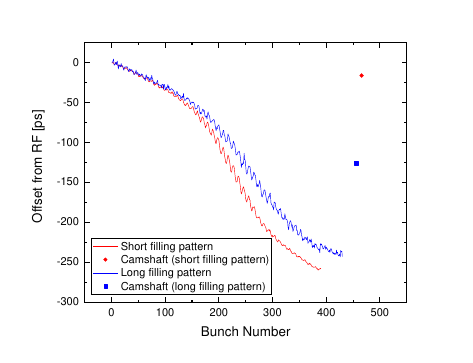}
    \caption{Shift of the center of the electron bunch with respect to its nominal position defined by the master clock. Measurements performed for two different filling patterns of the SLS. Image from \cite{Finizio2020}.}
    \label{fig:phaseShift}
\end{figure}

However, modern synchrotron lightsources are equipped with one or more superconducting passive cavities, tuned to the third harmonic of the master clock frequency. Those cavities, called \textit{third harmonic cavities}, are used to store a higher charge in the filling pattern, allowing one to reach higher electron currents and therefore a higher photon flux delivered to the beamlines \cite{Bosch1993}. However, third harmonic cavities affect the time structure of the stored electron beam. Of relevance for time-resolved STXM is that the center of the electron bunches is not constant with respect to the master clock, as shown in Fig. \ref{fig:phaseShift}. As the value of $N$ in Eq. \eqref{eq:RF_FPGA} is selected to be coprime with the total number of bunches in the filling pattern, each of the $N$ phase points in the excitation is equally sampled by all electron bunches. Therefore, a different position of the electron bunch with respect to the master clock will appear, from the point of view of the measurement, as a widening of the X-ray pulse width, i.e. a worsening of the achievable time resolution if compared to the case shown in Fig. \ref{fig:TR_STXM:ToAErrors}.

The phase shift of the electron bunches caused by the third harmonic cavities is a measurable quantity and is used as a beam diagnostic parameter. Therefore, if the position in the filling pattern of the electron bunch that emitted the photon, which was detected were known, one should be able to ``rebin'' the recorded photon count by taking into account the phase shift of the electron bunch with respect to the master clock. The position of the electron bunch that emitted the detected photon can be easily determined by using the bunch marker clock centrally provided by the timing system of the synchrotron light source. This marker is synchronized to the revolution frequency and is used, with an appropriate delay, to identify the first electron bunch in the filling pattern. By measuring the number of master clock cycles between the bunch marker and the detected photon count, the electron bunch from which the photon was emitted can be determined. Then, a lookup table storing the values of the electron bunch phase shift with respect to the master clock is consulted and the actual phase of the excitation that is probed by the detected photon is determined. Subsequently, the bin corresponding to the probed phase position is increased by one. This setup is installed at the Maxymus STXM at Bessy II \cite{Weigand2026} and is gradually replacing the previous demultiplexer implementation described in section \ref{sec:demux}.

\subsection{Time-of-arrival}
\label{sec:ToA_implementation}

\begin{figure*}
 \includegraphics[width=0.9\textwidth]{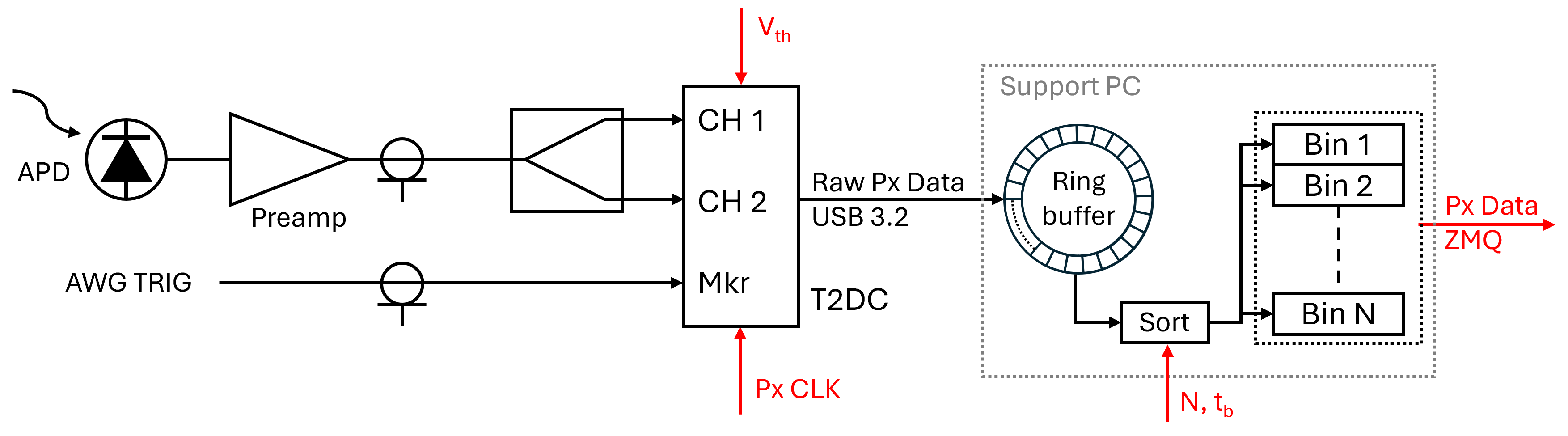}
 \caption{Simplified block diagram of the implementation of the time-of-arrival protocol. This setup utilizes a time-to-digital converter (T2DC) for the detection of the photon arrival time. A simulated CFD is performed by measuring the rising and falling edges of the APD pulse on two separate channels of the T2DC. A support computer is then used for binning and sorting the recorded arrival times.}
 \label{fig:ToA_concept}
\end{figure*}

The time-of-arrival detection protocol described in section \ref{sec:ToA} was implemented at the end of the 2010s at the PolLux endstation of the SLS \cite{Finizio2020}. The simplified block diagram of its implementation is shown in Fig. \ref{fig:ToA_concept}. The core component of this setup is the time-to-digital converter (T2DC), which records a high accuracy ($< 10$ ps resolution) timestamp each time that an input crosses a user-defined threshold. In the case of PolLux, the T2DC is a QuTAG device from the company QuTools GmbH \cite{Finizio2020}. 

\begin{figure*}[h]
\centering
\includegraphics[width=0.8\textwidth]{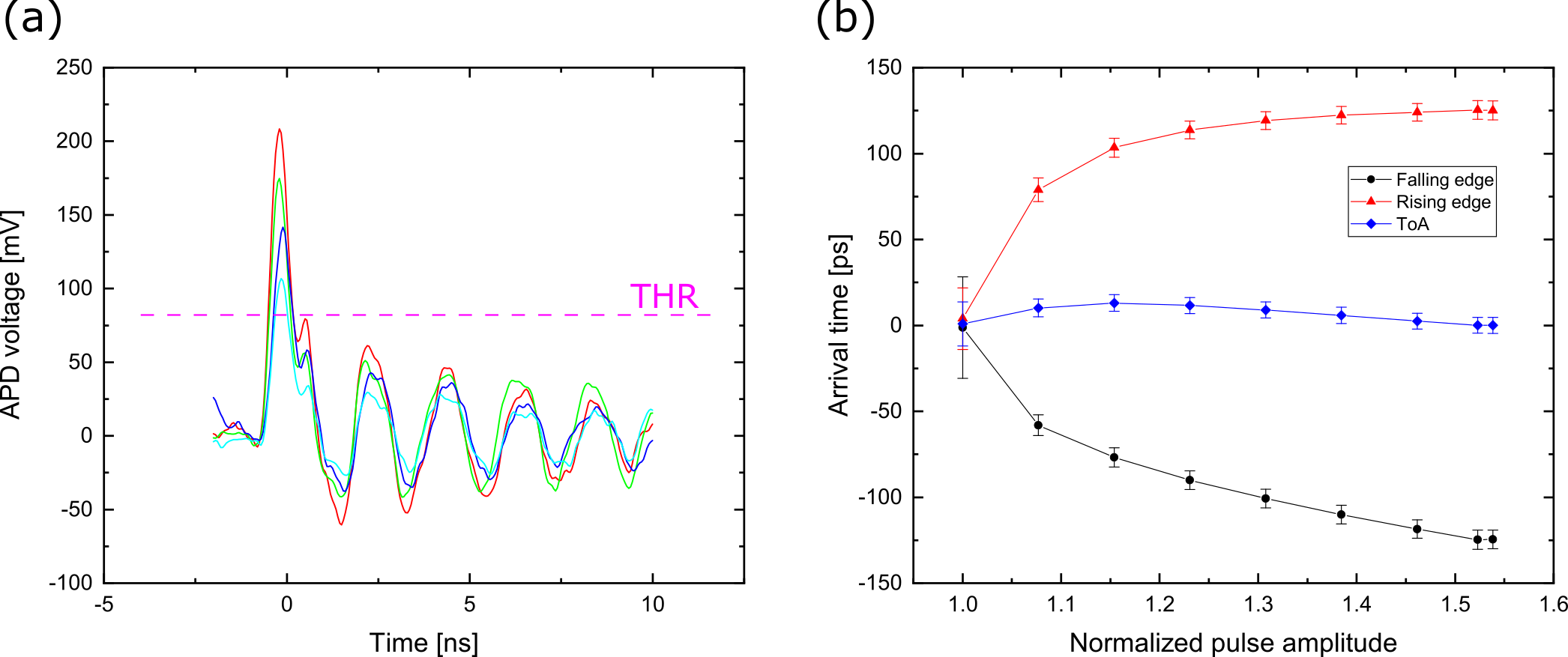}
\caption{Principle of the time-of-arrival detection protocol, illustrating the necessity to perform CFD detection. (a) Recorded oscilloscope traces from the APD of the PolLux beamline upon photon detection. Despite the X-ray photons having the same energy, a distribution of APD pulse heights is recorded, leading to a spread of recorded arrival times if the time at which the crossing of a defined voltage threshold (THR in the figure) is measured. (b) Arrival time error as a function of the APD pulse height if only the rising/falling edges of the pulse are recorded. If CFD is performed, the influence of the APD pulse height on the recorded arrival time can be almost completely eliminated. Image from \cite{Finizio2020}.}
\label{fig:ToA_APD}
\end{figure*}

The easiest implementation of the T2DC would be to use it to detect the rising edge of the ADC pulse and record the corresponding timestamp. However, the interaction between X-ray photons and the APD does not always occur in the same way. The X-ray photon can be absorbed in different parts of the junction, which leads to the generation of APD pulses with a distribution of different amplitudes. An example of some typical APD pulses is shown in Fig. \ref{fig:ToA_APD}(a). The variable amplitude of APD pulses leads, if one measures the time at which the pulse crosses a user-defined threshold, to an error in the photon arrival time and hence to a worsening of the achievable time resolution. This effect is shown in Fig. \ref{fig:ToA_APD}(b): if only the rising edge is measured, a pulse-height dependent error on the arrival time is introduced. This issue is already known in particle physics, for which a detection protocol known as \textit{constant fraction discrimination} (CFD) was developed \cite{Gedcke1967}. Thus, a ``software'' CFD detection was implemented by measuring when the rising and falling edges of the APD pulse cross a user-defined threshold, and calculating the photon arrival time as the average between the rising and falling edge timestamps. With this arrangement, the amplitude-dependent error on the recorded arrival time can be reduced to less than 10 ps, as shown in Fig. \ref{fig:ToA_APD}(b) \cite{Finizio2020}. The practical implementation of the ``software'' CFD is by using two channels of the T2DC (connected to the APD by means of a power splitter and cables of equal length), one set to measure the rising edge, and the second the falling edge. 

The third channel of the T2DC is used to monitor a marker synchronized with the excitation signal, usually generated with an AWG. The acquisition is controlled by the pixel clock, enabling the measurement of the timestamps when it is active. In normal operation conditions, the calculation of the average between the rising and falling edges and the calculation of the time difference between the recorded arrival time and the last marker is performed onboard the T2DC. The recorded time differences are then transmitted via a high bandwidth USB 3.2 connection to a support computer which stores them as they arrive in a ring buffer. They are then processed in a parallel thread and sorted into bins according to a user-defined bin width/number of bins. The binned timetrace is then transmitted via a zero message queue (ZMQ) interface to the STXM control software when the pixel clock is set to inactive \cite{Finizio2020}.

As discussed in section \ref{sec:ToA}, this setup does not require the excitation to be synchronized to a fractional multiple of the master clock, allowing for the excitation of dynamical processes at arbitrary frequencies. This also lifts another implicit limitation of the frequency locked setups, given by the limited number of independent counters that can be programmed into the FPGA, which limits the maximum duration of an excitation signal to $N/f_\mathrm{master}$. With the time-of-arrival detection setup presented here, in principle arbitrarily long signals can be investigated. Of course, a compromise between signal length, desired time step (bin width) and desired imaging statistics still needs to be taken into consideration when designing an experiment, as this setup does not change the total number of photons delivered to the sample.

As mentioned in section \ref{sec:ToA}, the time-of-arrival setup should also allow the attainment temporal resolutions below the width of the electron bunches, as the actual time of arrival of the photon is measured. In practice, however, the achieved temporal resolution with the setup installed at the PolLux endstation under a regular multibunch filling pattern beamtime has been of about 20-30 ps, determined by comparing streak camera measurements of the filling pattern with the data recorded with this method \cite{Finizio2020}. While this improves the achievable temporal resolution of about $5 \times$ with respect to the frequency-locked method described in section \ref{sec:RAM}, it still falls short of the temporal resolutions that would be achieved with low-$\alpha$ optics. A reason for the still relatively high uncertainty in the photon arrival time is given by the APD: depending on where the photon interacts within the junction, a delay in the rising time of the APD pulse could be introduced \cite{Baron2005}. A possible way to further improve the achievable time resolution is to move towards the novel low-gain avalanche diode (LGAD) technology \cite{Li2026}.

\subsection{4D Imaging}
\label{sec:4D}

The vast majority of time-resolved STXM imaging has been performed on two-dimensional (2D) systems in transmission geometry, i.e., the recorded images show the state of the sample averaged across its thickness. Although many magnetic systems can be accurately described by a 2D model, allowing an additional degree of freedom by introducing the third spatial dimension can open up a rich ensemble of dynamical behavior and novel functionalities \cite{FernandezPacheco2017}, which have recently attracted significant interest from the spintronic community. Although it is somewhat possible to infer the three-dimensional (3D) dynamical behavior of a system from 2D time-resolved imaging \cite{Butcher2025}, the direct visualization of the time-resolved 3D process, also called \textit{4D imaging}, is still preferable.

Spurred by this community-driven interest, significant progress has been made in the last decade towards the 3D imaging of magnetic systems, especially at the hard X-ray energies \cite{Donnelly2020}. This progress exploited techniques such as magnetic tomography \cite{Donnelly2017} and laminography \cite{Donnelly2020}. In both of those techniques, several 2D images of the sample, called projections, are acquired at different orientations of the sample with respect to the probing beam. The full set of projections is then used to computationally retrieve a 3D image of the sample \cite{Holler2019}. If combined with pump-probe imaging, a 4D image of the dynamical process can then be retrieved \cite{Donnelly2020}.

However, in the hard X-ray energy range, as ptychography \cite{Holler2019, Donnelly2020} is predominantly used for the acquisition of the single projections (see section \ref{sec:ptychoTR} for more details), the only accessible frequencies for 4D imaging at hard X-ray energies are integer multiples of the synchrotron master clock frequency \cite{Donnelly2020}. This is due to the requirement to use a 2D X-ray detector for ptychography, without the bandwidth necessary for the multi-probe imaging protocol described in section \ref{sec:freqLock}. In the soft X-ray energy range, instead, STXM is one of the standard imaging protocols for which multi-probe time-resolved imaging not limited to integer multiples of the master clock is routinely performed. 

To allow for time-resolved STXM imaging of 3D dynamical processes, the laminography protocol was extended to STXM with soft X-ray energies \cite{Finizio2022b, Witte2020}. In order to accommodate for the mechanical constraints of STXM imaging at soft X-ray energies (namely, the short focal distance of the zoneplate focusing optics, and short sample-detector distance) whilst still guaranteeing a sufficiently good spatial resolution, a laminography angle of 45$^\circ$ combined with a 35 nm outermost zone zoneplate was utilized, providing an achievable voxel size of 35-40 nm \cite{Finizio2022b, Witte2020}. 

\begin{figure}
 \includegraphics[width=0.4\textwidth]{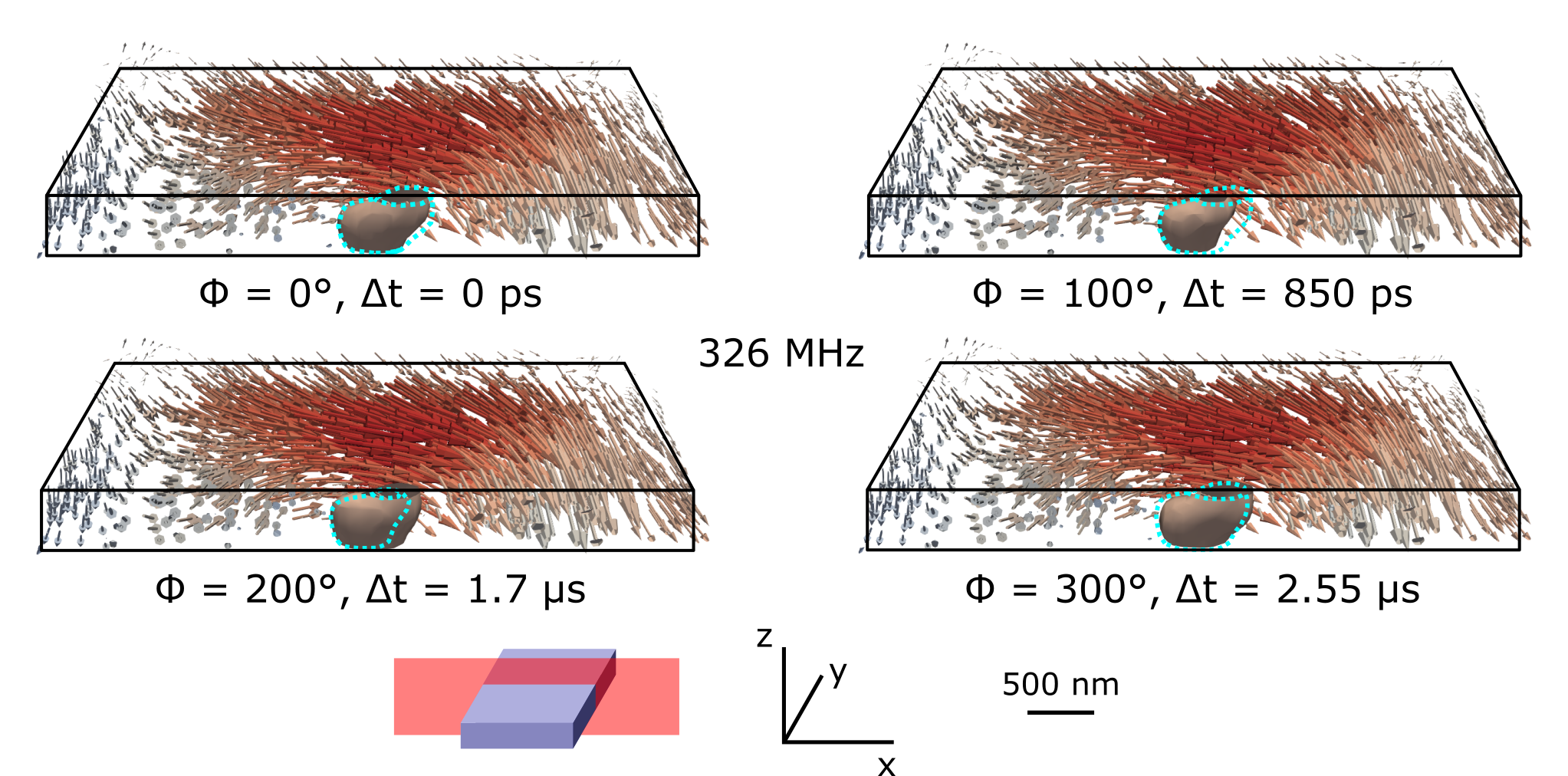}
 \caption{Snapshots of a 4D STXM-laminography image of the gyration dynamics of a magnetic vortex in a 150 nm thick Co$_{40}$Fe$_{40}$B$_{20}$ microstructured square. The excitation was performed by applying an in-plane magnetic field at a frequency of about 326 MHz. The vortex core, where the magnetization exhibits an out of plane orientation, is marked by the brown surface. The position of the vortex core at the start of the excitation is marked by the cyan dashed line. Further details in \cite{Finizio2022b}.}
 \label{fig:TR_lamni_vortex}
\end{figure}

A proof-of-concept time-resolved STXM laminography experiment using a 150 nm thick Co$_{40}$Fe$_{40}$B$_{20}$ microstructured element stabilizing a Landau flux closure pattern \cite{Hubert1998}, was performed \cite{Finizio2022b}. Hereby, for each of the 50 projections used to reconstruct the magnetic 3D configuration of the sample, a $N=23$ frame time-resolved STXM image was acquired for each of the two frequencies investigated (326 MHz, corresponding to $M=15$ and 913 MHz, corresponding to $M=42$ with the notations described in section \ref{sec:freqLock}). The reconstruction of the 3D time-resolved image was then carried out by performing a separate 3D reconstruction for each of the frames in the time-resolved STXM image (in the work presented in \cite{Finizio2022b}, the 23 frames were binned in 7 final frames to improve the statistics of the single frames that were then used for the 3D reconstruction) \cite{Finizio2022b}. A rendering of some of the frames of the reconstructed 3D time-resolved image at 326 MHz is shown in Fig. \ref{fig:TR_lamni_vortex}, where the combined motion and deformation of the out-of-plane magnetized vortex core at the center of the microstructure can be observed. The reliability of the 3D time-resolved reconstruction was verified by comparing the experimental data with micromagnetic simulations \cite{Finizio2022b}.

\begin{figure*}
    \includegraphics[width=0.8\textwidth]{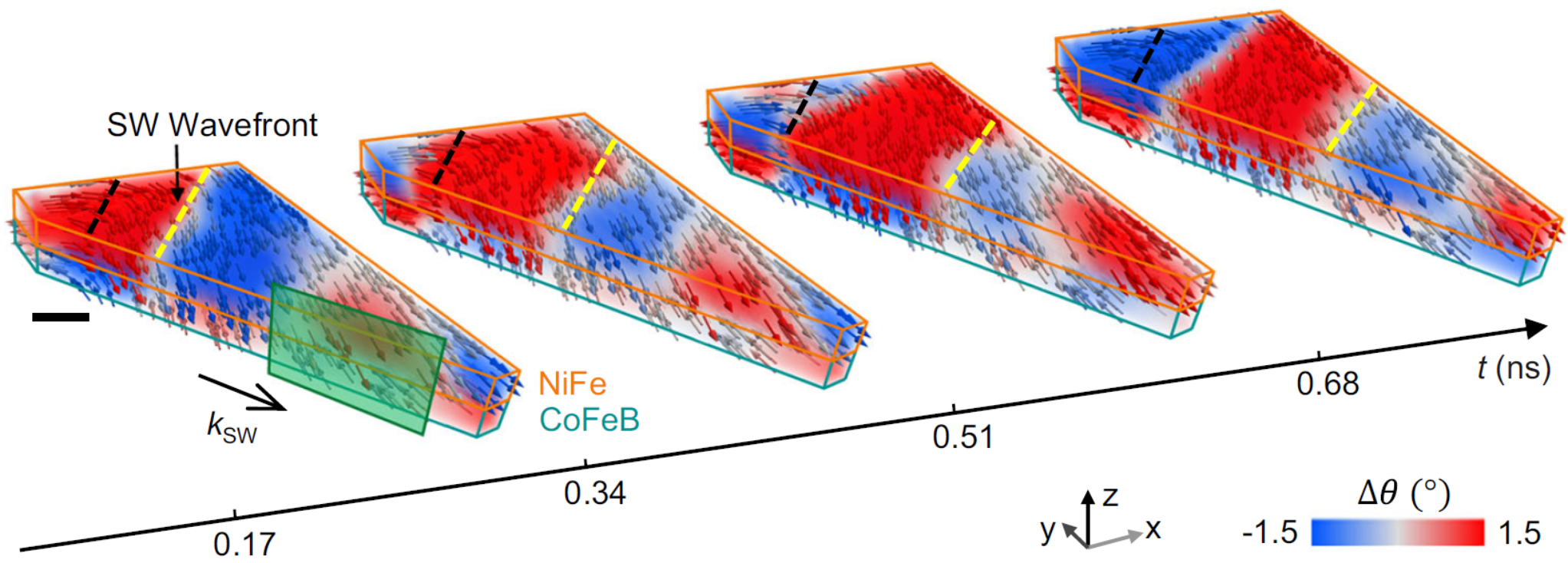}
    \caption{Snapshots of a time-resolved 3D STXM-laminography image showing the propagation of planar spinwave wavefronts emitted by a domain wall in a SAF microstructured element. Image from \cite{Girardi2024}.}
    \label{fig:TR_lamni_spinwaves}
\end{figure*}

Beyond the proof-of-concept experiment described in \cite{Finizio2022b}, time-resolved 3D STXM imaging has been utilized for investigating the 3D dynamical behavior of magnons in a synthetic antiferromagnetic (SAF) material \cite{Girardi2024}. In this study, 3D imaging was necessary to visualize the localization of the Damon-Eshbach magnons \cite{Eshbach1960} at the interface between the two layers of the SAF, and to visualize the depth dependence of the interference pattern of magnons generated by a curved domain wall. A summary of the results is shown in Fig. \ref{fig:TR_lamni_spinwaves}, underlining how 4D imaging is necessary to unravel the complex dynamical patterns that occur in a 3D system.

Due to the requirement of acquiring several projections (for the works described in \cite{Finizio2022b, Girardi2024}, 37 to 50 projections were acquired for each TR image) combined with the necessity to have a pixel dwell time sufficiently high as to guarantee good statistics for each frame of the time-resolved images (typically around 100-200 ms/px for the photon flux of the PolLux beamline \cite{Finizio2022b, Girardi2024}), 3D time-resolved STXM imaging is a time intensive technique, with one typical 3D time-resolved image requiring about 2--3 days of continuous measurements at the PolLux beamline \cite{Finizio2022b}. This implies that, at least presently (see section \ref{sec:future} for additional discussions on this point), the throughput of 3D time-resolved STXM imaging experiments is low, and requires samples exhibiting dynamical processes that are stable over the course of several days of continuous excitation. Additionally, particular attention must be paid to the quality of the vacuum inside the STXM experimental chamber, as X-ray induced carbon deposition on the sample should be minimized in order to avoid unwanted artifacts in the reconstruction of the 3D images \cite{Witte2020, Finizio2022b, Watts_C_cleaning}. Nonetheless, this method is one of the few experimental techniques allowing for the full 4D imaging of dynamical processes at the nanometric and sub-nanosecond scale.

\subsection{Imaging of auto-oscillatory dynamics}

All the detection protocols described above require synchronization with the applied excitation, either through the master clock of the synchrotron or by providing a marker synchronized with the excitation, as this is used for the sorting of the recorded photon counts in the correct time bin \cite{Weigand2022, art:puzic_TR_STXM, Finizio2020}. This requirement, however, limits the ensemble of dynamical processes that can be investigated by the technique. Non-locked dynamics, i.e. periodic processes occurring at a frequency different from the one used to externally excite the system (an example could be the Bose-Einstein condensation of parametrically excited magnons \cite{Demokritov2006}), or auto-oscillatory processes, i.e. systems oscillating at an intrinsic frequency (e.g. spin-torque vortex oscillators, whose oscillation frequency can be controlled by a direct current signal \cite{Locatelli2014}) cannot be imaged with the techniques described above, as the oscillations are not locked to the external excitation signal. However, in such a system, the X-ray absorption cross section still exhibits an oscillatory behavior at the same frequency of the dynamical process, implying that the recorded photon count rate in a STXM experiment shows an oscillatory behavior at that frequency, which could in principle be inferred from the absolute recorded photon arrival times. 

Therefore, by utilizing the time of arrival measurement setup described above \cite{Finizio2020} to record the absolute photon arrival times, one should be able to also resolve non-locked dynamical processes. However, the typical photon rates of a STXM beamline are on the order of 10$^6$-10$^7$ ph/s \cite{pollux_STXM} and the typical frequencies of magnetodynamical processes in ferroic systems are in the 10 MHz--100 GHz range \cite{Marrows2016}, such a dataset would be sparse, requiring specific methods for the reconstruction of the oscillating information. In particular, it is possible to reconstruct the power spectrum of an oscillating system with a sparsely sampled dataset through the use of the Schuster periodogram \cite{Lomb1976, Scargle1982, Schuster1898}, which was originally developed for the study of astronomical objects with oscillating magnitude \cite{Schuster1898}. 

\begin{figure}
  \includegraphics[width=0.4\textwidth]{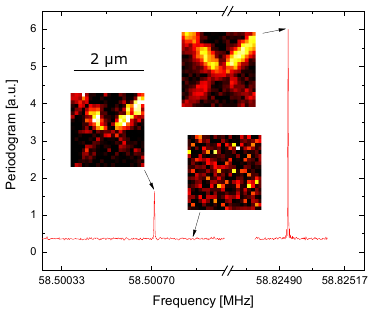}
 \caption{Example of a power spectrum calculated according to Eq. \eqref{eq:periodogram} of a sample that was excited with two RF signals separated by approximately 300 kHz, the value of which was not known to the detection setup. The periodogram reconstruction allows one to identify the excitation frequencies without prior knowledge on their values. Image adapted \cite{Finizio2022}.}
 \label{fig:Schuster}
\end{figure}

In a simplified description, the Schuster periodogram is a brute force calculation of several discrete Fourier transforms at different frequencies, allowing one to determine the power spectrum for each given frequency $f$ in the signal \cite{Lomb1976, Scargle1982, Schuster1898}:

\begin{equation}
 P(f; t_n) = \left| \sum_{n=1}^{N} \mathrm{e}^{-2\pi i f t_n}\right|^2,
 \label{eq:periodogram}
\end{equation}

where the dataset consists of $N$ timestamps $t_n$. An example of a power spectrum calculated from time of arrival photon data using the Schuster periodogram is shown in Fig. \ref{fig:Schuster}, where the peaks in the power spectrum identify dynamical processes occurring within the sample \cite{Finizio2022}. Combined with STXM, this technique allows the localization of the regions of the sample oscillating at the given frequency \cite{Finizio2022}. It should also be noted that the usual Nyquist frequency criterion for uniform sampling does not apply, allowing for the identification of frequency signals at much higher frequencies than the photon rate. Instead, the limit on the resolvable frequencies is determined by the precision of the photon arrival timestamp, given by $f_\mathrm{max} = 10^D/2$, where $D$ is the number of decimal places in the timestamp \cite{VanDerPlas2018}. For the case described in \cite{Finizio2022}, the resolution of the timestamping is of 0.5 ps, leading to a theoretically maximum identifiable frequency of 500 GHz. In practice, the actually detectable frequencies will be similar to those achievable by the time-of-arrival setup described above in section \ref{sec:ToA_implementation}.

If the entire frequency range of 500 GHz were to be explored, the number of periodograms that would have to be investigated is on the order of 10$^{10}$-10$^{11}$, which is computationally intractable \cite{Finizio2022}. However, by combining graphics processing unit (GPU) assisted computation, subdivisions of the acquired data into smaller datasets, and some preliminary knowledge (e.g., from micromagnetic simulations or other preliminary measurements), the calculation of the power spectrum can be performed in acceptable timeframes \cite{Finizio2022}. The further evolution of GPU-assisted computing in the recent years will allow one to expand even further the frequency range that can be ``blindly'' investigated.

Through this method, the ensemble of dynamical processes that can be investigated by TR-STXM imaging was extended to non-locked dynamics.

\section{Implementations - The future}
\label{sec:future}

This section provides an outlook of possible future developments for the time-resolved STXM imaging technique, and how they can be integrated with the currently ongoing upgrades of synchrotron lightsources in the world to diffraction limited storage rings (DLSRs). Novel implementations for APD detectors, based on the relatively new thin entrance window low gain avalanche diode (LGAD) technology \cite{Zhang2022} will also be reviewed here. Finally, a review of the current options for time-resolved ptychography will be discussed.

\subsection{Fourth generation lightsources - managing detector pileup}
\label{sec:DLSR}

Diffraction-limited synchrotron (DLSR) light sources, or 4$^\mathrm{th}$ generation light sources, are a novel storage ring design that delivers a significant increase in the coherent photon flux delivered to the beamlines compared to 3$^\mathrm{rd}$ generation storage ring design \cite{Raimondi2023}. At the time of writing, several synchrotron lightsources around the world have undergone the upgrade to a DLSR, and several more upgrades are planned or undergoing. For STXM imaging, an increase in the coherent photon flux delivered to the instrument will enable faster acquisition of images and routine acquisition of high-resolution images. The same simplified reasoning could be applied for time-resolved imaging: more light means faster/better images.

\begin{figure}
    \centering
    \includegraphics[width=0.45\textwidth]{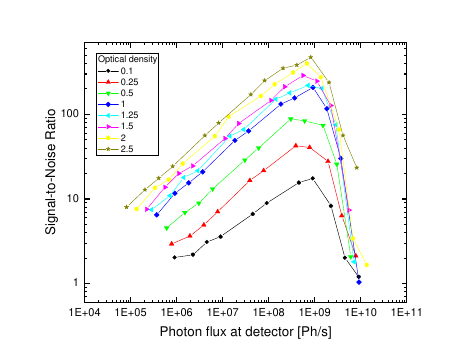}
    \caption{Simulated effect of detector pileup on the signal to noise ratio for different optical densities (i.e., thickness) of the sample. The simulations assume a master clock of 500 MHz. A significant drop in the signal to noise ratio can be observed for a photon flux of about 10$^9$ Ph/s on the detector, where the pileup probability is about 50\%. Image from \cite{Finizio2021}.}
    \label{fig:pileup}
\end{figure}

However, a higher photon flux is only half the equation, both in the space and time domains. In the space domain, the efficient control of sample vibrations and imaging overheads is being tackled in order to improve the imaging throughput and efficiently utilize all the light that is delivered to the endstation \cite{Finizio2026}. In the time domain, there are two main considerations: the first is that the DLSR design is in general not compatible with low-$\alpha$ optics, meaning that methods such as the time-of-arrival detection described in section \ref{sec:ToA} need to be utilized to achieve temporal resolutions better than about 70-100 ps. The second consideration is that the increase in the delivered photon flux will lead to an increased probability of detector pile-up events. As the detection methods described in section \ref{sec:pumpProbe} and their implementations described in section \ref{sec:present} assume that either \textit{zero} or \textit{exactly one} X-ray photon can be detected for each electron bunch, pileup events lead to an erroneous counting of the detected photons and to a significant decrease in the measured signal-to-noise ratios in the time-resolved images, as depicted in Fig. \ref{fig:pileup} \cite{Finizio2021}.

As the amplitude of the APD pulses depends on where the photon interacts within the diode junction, threshold-based methods for pileup management/rejection cannot be reliably employed, and for high photon fluxes, rejecting multi-photon events would lead to an inefficient usage of the delivered light \cite{Finizio2021}. One option to overcome this issue could be to replace the APD with a multi-sector diode, reducing the probability of a multi-photon event in the same sector of the diode and allowing one to use the same methods presented in section \ref{sec:present} for higher photon fluxes, albeit at the price of a higher complexity in both the detector hardware (each diode sector would act as an independent APD detector, meaning a separate preamplifier circuit for each diode sector) and in the signal processing hardware (each implementation needs to be repeated for each diode sector, as they are treated as independent detectors). The development of solutions to manage multi-channel detection is therefore encouraged, perhaps by exploring multi-channel detection mechanisms used in particle physics (e.g., for detection of muons \cite{Lecoq2020}).

\subsection{Low Gain Avalanche Diodes}
\label{sec:LGAD}

At the time of writing, APDs are the only suitable detector for time-resolved STXM imaging, due to the requirement for high bandwidths (the typical APD used for time-resolved STXM has a bandwidth of 900 MHz-1 GHz). While APDs have been of great use for time-resolved STXM imaging, they present a major limitation in the energy range that can be detected. In particular, due to a relatively thick entrance window, commercial APD detectors only exhibit meaningful quantum efficiencies for photon energies above approximately 600 eV \cite{Baron2005, VanWaeyenberge2006, Acremann2007}, limiting the ensemble of systems that can be investigated to those with elemental edges at energies above 600 eV. Notable exclusions are the K absorption edges of Carbon (ca. 280 eV), Nitrogen (ca. 390 eV), and Oxygen (ca. 530 eV) and the L absorption edges of Titanium (ca. 430 eV). Access to those elemental edges allows investigating the dynamics of ferroelectric perovskite oxides such as BaTiO$_3$ by investigating the crystal field split hybridized O 2p orbitals using the X-ray linear dichroism effect as contrast mechanism to visualize ferroelectric domain configurations \cite{Butcher2025b, Butcher2024}.

An APD detector combining a bandwidth above the master clock frequency (ideally at least twice the master clock frequency) with satisfactory quantum efficiency below 600 eV would allow for the unlocking of a large ensemble of dynamical processes. A promising avenue for this is a new concept for avalanche diodes, called low-gain avalanche diode (LGAD). LGADs have quickly risen in popularity as an alternative to APDs thanks to the combination of a uniform and well-controllable gain layer with excellent timing resolution and, especially, thin ($< 300$ nm) entrance windows \cite{Fernandez2015, Zhang2022}. Thin window LGADs have been successfully used for ptychography imaging down to 530 eV \cite{Butcher2025b}, and dedicated research aims to further improve the quantum efficiency at low energies \cite{Zhang2022}. If these efforts bear fruit, ultrathin window LGADs could also find applications for time-resolved STXM and allow investigation of dynamical processes beyond those occurring in 3d transition metal ferromagnets and rare-earth containing systems.

\subsection{Time-resolved ptychography}
\label{sec:ptychoTR}

As briefly mentioned in section \ref{sec:STXM}, the achievable spatial resolution of STXM imaging is limited by the beam spot size that can be achieved with the focusing optics. The current limit, governed by increasingly challenging lithographical fabrication, is about 7 nm \cite{Roesner2020}. One pathway to improve the spatial resolution beyond the limits of the focusing optics is to perform CDI, as the spatial resolution for this is ultimately determined by the solid angle that can be detected downstream of the sample in far-field conditions \cite{Guizar-Sicairos2008, Fienup1982}, i.e. by the size of the detector and the wavelength of the X-rays. In CDI, the coherent diffraction pattern caused by the interaction of a coherent X-ray beam with the sample is recorded by a 2D X-ray detector in far field imaging conditions. In a simplified description, imaging at the far field would correspond to acquiring the Fourier transform of the object \cite{Nugent2009}, i.e. the original real-space image of the object could be retrieved by calculating the inverse Fourier transform of the scattering pattern. However, as X-ray detectors only detect the intensity of the light that illuminates them, the phase information is lost, and one cannot retrieve the real-space image of the object simply by calculating the inverse Fourier transform \cite{Nugent2009, Fienup1982, Guizar-Sicairos2008}. Therefore, the recovery of the phase information is one of the core components of CDI, and over the years several methods have been developed to assist with phase recovery. One example is X-ray holography \cite{Eisebitt2004}, where the phase information is encoded in the scattering pattern by fabricating small reference holes adjacent to the sample and recording the interference pattern between the two.

Another approach is to perform the phase recovery through an iterative computation \cite{Fienup1982}. This iterative computation is performed by starting with a random guess of the real-space image of the object and calculating its Fourier transform. In Fourier space, the magnitude of the Fourier transform is set to the square modulus of the amplitude measured by the 2D detector, but the phase is left unchanged. This is the Fourier constraint of the phase recovery algorithm. Then, the inverse Fourier transformation is calculated, and a real-space constraint is applied to the calculated real-space image (e.g., in classical CDI, the real space constraint is to set the regions surrounding the sample - fabricated at the center of a non-transparent region - to non-transparent). Then, the Fourier transform is calculated again, and the steps described above are iteratively repeated, until a real-space image of the object is recovered with an acceptable quality.

One limitation of ``classical'' CDI is that the object is fabricated at the center of a non-transparent region of the sample, i.e. lithographical fabrication is usually required, and the regions of interest in the sample are limited in size \cite{Eisebitt2004}. In recent years, a CDI technique combining its superior resolution with the possibility to perform arbitrarily large scans without the requirement for lithographical fabrication has undergone rapid development. This technique is called X-ray ptychography imaging and combines CDI with scanning microscopy \cite{Hoppe1969, pfeiffer_2018}. In ptychography, a dataset is acquired by recording 2D diffraction patterns at different positions of the sample, with the caveat that the illumination spot size partially overlaps with the one from the neighboring points in the scan. This allows introduction of an overlap constraint in the real-space part of the phase recovery algorithm. As the reconstruction of a ptychographic image requires heavy computational resources, this method, albeit originally proposed in 1969 \cite{Hoppe1969}, has been only recently used for experimental investigations thanks to the significant improvement in computational power. With this method, spatial resolutions routinely below 10 nm at soft X-ray energies could be achieved \cite{Butcher2025c, Shapiro2020, Harrison2025}.

In terms of achievable spatial resolutions, ptychography sets the state-of-the-art in soft X-ray scanning microscopy. It would therefore be desirable to combine the technique with time-resolved imaging to be able to acquire high spatial and temporal resolution images of dynamical processes. However, the requirement to use a 2D X-ray detector for the acquisition of the diffraction patterns is a hindrance for time-resolved imaging. At the time of writing, no 2D soft X-ray detector able to either measure the photon arrival time or to distinguish photons emitted from neighboring electron bunches exists, hindering the possibility to perform measurements using the single pump-multiple probe methods described in sections \ref{sec:freqLock} and \ref{sec:ToA}. Even performing the classical pump-probe approach with a SB filling pattern, or in hybrid mode (some detectors, such as e.g. the upcoming Matterhorn detector \cite{Mezza2023}, can be gated sufficiently fast to allow operation with the camshaft bunch), is challenging, as ptychography may lead due to unfeasible acquisition times for a full time-resolved scan.

A workaround to allow for the acquisition of \textit{some} dynamical processes with ptychographic imaging is to perform single pump-multiple probe imaging with the excitation locked to an integer multiple of the master clock. This allows, as described in section \ref{sec:classicPP}, probing the same phase point of the excitation by all of the electron bunches. The temporal resolution for this is reduced due to the phase shift caused by the third harmonic cavities in the ring - see section \ref{sec:phaseShift} for more details. Time-resolved ptychography imaging both at the hard \cite{Donnelly2020} and soft \cite{Butcher2025} X-ray energies has been performed in this way. 

Unlocking superior spatial resolutions for time-resolved imaging with ptychography outside of the ``easy'' case of multiples of the master clock frequency hinges on detector technology. Some progress towards the ability to detect the photon arrival time, e.g., in the TimePix project \cite{Llopart2022}, is being made, but a future time-resolved detector needs not only to be able to provide an arrival timestamp for the photon, but also to manage the other requirements (i.e., high dynamic range, ability to detect high photon fluxes, and sensitivity down to the soft X-ray energies) for soft X-ray ptychography imaging. An alternative, but at the price of reduced experimental throughput, is to perform the measurements with a gated detector (e.g., the upcoming Matterhorn detector \cite{Mezza2023}), using only the camshaft bunch. Here, an experiment combining regular time-resolved STXM for high-throughput imaging and time-resolved ptychography for the high resolution imaging of a selected subset of experimental data could be envisioned as a possible compromise solution.

\section{Conclusions}

Scanning transmission X-ray microscopy is a powerful and flexible technique for the imaging of ferroic systems thanks to the possibility to combine nanometric spatial resolutions with several contrast mechanisms offered by the use of monochromatic soft X-ray radiation of different polarization. One of the main strengths of STXM imaging, however, lies in the fact that a broadband APD can be utilized as photon detector, allowing for the possibility to acquire time-resolved images through the pump-probe protocol. This review commenced with an overview of the operating principles of pump-probe imaging and the variations to the classical pump-probe protocol that are enabled in time-resolved STXM imaging. These protocols, which are called \textit{single pump-multiple probe}, allow for the imaging of dynamical processes with a widely selectable comb of frequencies (and even a freely selectable frequencies for the \textit{time-of-arrival} protocol) using the entire synchrotron filling pattern, i.e. without sacrificing on imaging statistics. The different practical implementations of the single pump-multiple probe protocol have also been reviewed, ranging from the first implementations in the early 2000s at the ALS \cite{Acremann2006, VanWaeyenberge2006} to the time-of-arrival setup currently installed at the SLS \cite{Finizio2020}. 

In addition, an overview of the possible future implementations of the single pump-multiple probe protocol in STXM imaging was given. In particular, the opportunities and associated challenges of the recently ongoing upgrade of synchrotrons to DLSRs for time-resolved STXM was reviewed, with some possible solutions to face the increase in pileup events, e.g. through the use of multichannel detectors. Finally, the current status of time-resolved soft X-ray ptychography imaging was described, which recently revolutionized the achievable spatial resolutions in soft X-ray microscopy, and the challenges in implementing the single pump-multiple probe protocol for this imaging technique.

Thanks to the upgrade to DLSRs and in time resolved imaging protocols, bright days are ahead of us, which have the potential to usher in another exciting 20 years of time-resolved STXM for the investigation of dynamical processes, not only limited to condensed matter physics.

\begin{acknowledgments}
The authors thank A. Bergamaschi and J. Zhang from the PSI detector group for helpful discussions on LGAD detectors and A. Doll from the \textmu SR at PSI for discussions about multi-channel timing detection methods. T.A.B. acknowledges funding from the European Regional Development Fund (ERDF).
\end{acknowledgments}

\bibliography{bibliography}

\end{document}